\documentclass[manuscript,screen]{acmart}
\usepackage{multirow} 

\newcommand{\modified}[1]{\textcolor{blue}{#1}} 

\begin{document}

\title{A Survey of State-of-the-Art on Blockchains: Theories, Modelings, and Tools}

\author{Huawei~Huang, Wei~Kong, Sicong~Zhou, Zibin~Zheng}
\authornote{Both authors are corresponding authors.}
\orcid{0000-0002-7035-6446}
\affiliation{%
	\institution{Sun Yat-Sen University}
	\department{School of Data and Computer Science}
	\streetaddress{132 East Waihuan Road, Higher Education Mega Center}
	\city{Guangzhou}
	\state{Guangdong}
	\postcode{510006}
	\country{China}
	}
\email{huanghw28@mail.sysu.edu.cn; zhzibin@mail.sysu.edu.cn}


\author{Song~Guo}
\authornotemark[1]
\affiliation{%
	\institution{The Hong Kong Polytechnic University}
	\department{Department of Computing}
	\city{Hong Kong SAR}
	\country{China}
	}
\email{song.guo@polyu.edu.hk}


\begin{abstract}

To draw a roadmap of current research activities of the blockchain community, we first conduct a brief overview of state-of-the-art blockchain surveys published in the recent 5 years. We found that those surveys are basically studying the blockchain-based applications, such as blockchain-assisted Internet of Things (IoT), business applications, security-enabled solutions, and many other applications in diverse fields. However, we think that a comprehensive survey towards the essentials of blockchains by exploiting the state-of-the-art theoretical modelings, analytic models, and useful experiment tools is still missing. To fill this gap, we perform a thorough survey by identifying and classifying the most recent high-quality research outputs that are closely related to the theoretical findings and essential mechanisms of blockchain systems and networks. Several promising open issues are also summarized finally for future research directions. We wish this survey can serve as a useful guideline for researchers, engineers, and educators about the cutting-edge development of blockchains in the perspectives of theories, modelings, and tools.
    
\end{abstract}



\maketitle


{\color{black}

\section{Introduction}\label{sec:introduction}

 \modified{Centralized security mechanisms are prone to Single Point of Failure, meaning that once a centralized component is compromised, the whole system would cease to function. The decentralization of blockchain can eliminate such concern without the need of a trusted third party. With the benefit of decentralized characteristics,} blockchains have been deeply diving into multiple applications that are closely related to every aspect of our daily life, such as cryptocurrencies, business applications, smart city, Internet-of-Things (IoT) applications, and etc.  
 In the following, before discussing the motivation of this survey, we first conduct a brief exposition of the state-of-the-art blockchain survey articles published in the recent few years.
 
 }


 \subsection{Taxonomy of State-of-the-art Blockchain Surveys}
 
\begin{figure*}[t]
\centering
	\vspace{-3mm}
\includegraphics[width=0.7\textwidth]{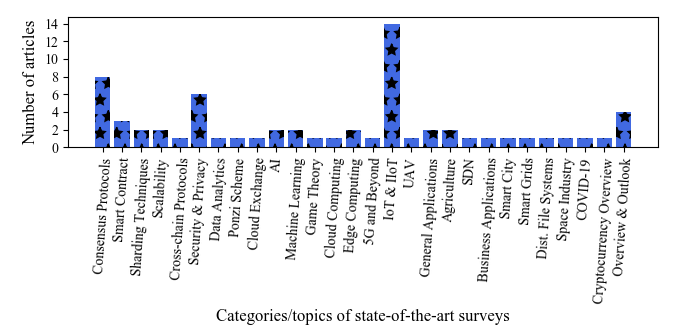}
	\vspace{-2mm}
\caption{The categories and number of state-of-the-art blockchain-related surveys published in recent few years.}
\label{fig:category}
	\vspace{-5mm}
\end{figure*}

\begin{figure*}[t]
\centering
	\vspace{-4mm}
\includegraphics[width=0.8\textwidth]{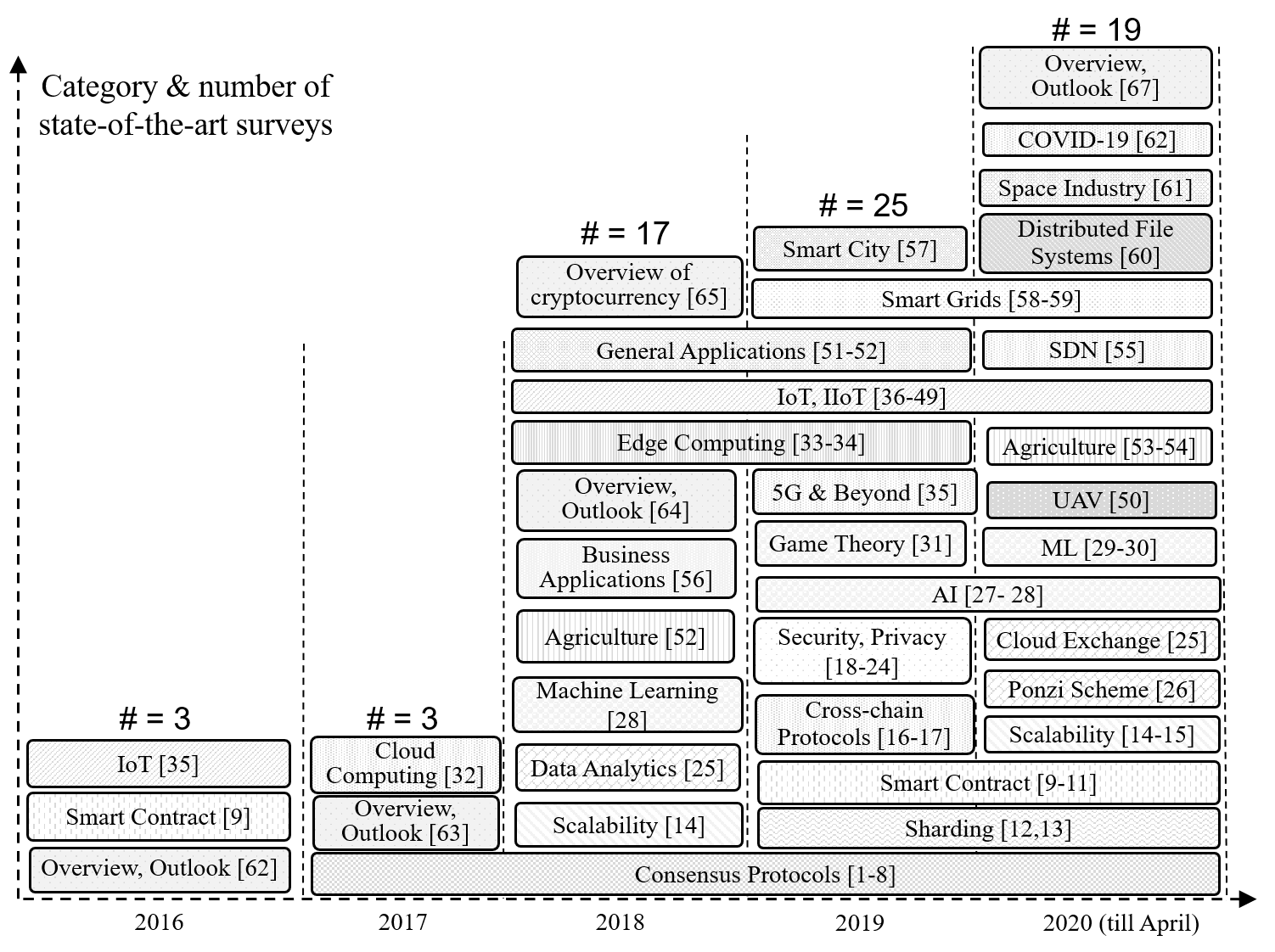}
	\vspace{-3mm}
\caption{\modified{The category and number of state-of-the-art blockchain surveys organized in a chronological order.}}
\label{fig:classification}
	\vspace{-3mm}
\end{figure*}

 To identify the position of our survey, we first collect 67 state-of-the-art blockchain-related survey articles. The numbers of each category of those surveys are shown in Fig. \ref{fig:category}. We see that the top-three popular topics of blockchain-related survey are IoT \& IIoT, Consensus Protocols, and Security \& privacy.
We also classify those existing surveys and their chronological distribution in Fig. \ref{fig:classification}, from which we discover that i) the number of surveys published in each year increases dramatically, and ii) the diversity of topics also becomes greater following the chronological order.
In detail, we summarize the publication years, topics, and other metadata of these surveys in Table \ref{Table:taxonomy} and Table \ref{Table:taxonomyPart2}. 
Basically, those surveys can be classified into the following 7 groups. \modified{The overall principal of the collection is based on different aspects of blockchain covered in the surveys. In \textit{Group-1},  different abstraction layers of blockchain protocols and intrinsic properties are the main focus. In \textit{Group-2},  the behaviour of blockchain's clients are analysed by means of data mining. In \textit{Group-3}, blockchain as a complicated and rewarding environment, with hard choices made by means of AI or game theory. Also, surveys that analysed the integration of blockchain and decision making techniques are also classified in this group. In \textit{Group-4}, the integration of blockchain and different communication techniques. In \textit{Group-5} and \textit{Group-6}, the applications of blockchain are reviewed. We singled out surveys on IoT applications due to the popularity. \textit{Group-7} are the works of holistic overview of blockchain.
}

\subsubsection{\textbf{Blockchain Essentials}}

The first group is related to the essentials of the blockchain. A large number of consensus protocols, algorithms, and mechanisms have been reviewed and summarized in \cite{sankar2017survey, yuan2018blockchain, wang2018survey, garay2018sok, nguyen2018survey, wang2019survey, bano2017consensus, xiao2020survey}.  For example, motivated by lack of a comprehensive literature review regarding the consensus protocols for blockchain networks,  Wang \textit{et al.} \cite{wang2018survey} emphasized on both the system design and the incentive mechanism behind those distributed blockchain consensus protocols such as Byzantine Fault Tolerant (BFT)-based protocols and Nakamoto protocols. From a game-theoretic viewpoint, the authors also studied how such consensus protocols affect the consensus participants in blockchain networks.

 During the surveys of smart contracts \cite{atzei2016survey, dwivedi2019blockchain, zheng2020overview}, Atzei \textit{et al.} \cite{atzei2016survey} paid their attention to the security vulnerabilities and programming pitfalls that could be incurred in Ethereum smart contracts.
Dwivedi \textit{et al.} \cite{dwivedi2019blockchain} performed a systematic taxonomy on smart-contract languages, while Zheng  \textit{et al.} \cite{zheng2020overview} conducted a survey on the challenges, recent technical advances and typical platforms of smart contracts.

 Sharding techniques are viewed as promising solutions to solving the scalability issue and low-performance  problems of blockchains. Several survey articles  \cite{wang2019sok, Yu2020AccessSharding} provide systematic reviews on sharding-based blockchain  techniques. For example,  Wang \textit{et al.} \cite{wang2019sok} focused on the general design flow and critical design challenges of sharding protocols.
 Next, Yu \textit{et al.} \cite{Yu2020AccessSharding} mainly discussed the intra-consensus security, atomicity of cross-shard transactions, and other advantages of sharding mechanisms.

Regarding scalability, Chen \textit{et al.} \cite{chen2018research} analyzed the scalability technologies in terms of efficiency-improving and function-extension of blockchains, while Zhou \textit{et al.} \cite{zhou2020solutions} compared and classified the existing scalability solutions in the perspective of different layers.
Then, Zamyatin \textit{et al.} \cite{zamyatin2019sok} conducted a systematic classification of protocols for cross-chain communication.
\modified{Further on interoperability, Belchior \textit{et al.} \cite{belchior2020survey} defined related terms and provided interesting directions.}
During the investigations  \cite{taylor2019systematic, dasgupta2019survey, ma2019survey, tariq2019security, feng2019survey, soni2019comprehensive} on  security \& privacy issues,
Taylor  \textit{et al.} \cite{taylor2019systematic} reviewed the cyber security space of blockchains including security of blockchain in different directions such as IoT, AI data and sidechain.
 Dasgupta \textit{et al.} \cite{dasgupta2019survey} discussed general security issues of blockchains from theory to implementation, such as vulnerability, malicious attacks, risks of blockchain applications, and etc.
 Ma \textit{et al.} \cite{ma2019survey} focused on security, privacy and trust issues in crowdsourcing services.
 Under the background of big data, Tariq \textit{et al.} \cite{tariq2019security} reviewed the security challenges of fog computing-enabled IoT applications, in which blockchain techniques are playing a role of security enabler.
 In contrast, \cite{feng2019survey, soni2019comprehensive} emphasized on the privacy issues of  blockchain systems and blockchain-based applications.

\begin{table*}[t]
\centering
\footnotesize
\caption{\modified{Taxonomy of existing blockchain-related surveys (Part 1).}}
\label{Table:taxonomy}
\begin{tabular}{|p{0.11\textwidth}|p{0.14\textwidth}|p{0.1\textwidth}|p{0.029\textwidth}|p{0.46\textwidth}|}%
\hline
 \textbf{Group}&\textbf{Category}&\textbf{Ref.}&\textbf{Year} &\textbf{Topic}\\
\hline


  \multirow{18}*{\textit{Group-1:}}
    &  \multirow{6}*{Consensus} & 
    Sankar  \cite{sankar2017survey} & 2017 &  Consensus protocols on blockchain applications\\

	\cline{3-5}
	
 \multirow{18}*{\textbf{Blockchain}}
   &   \multirow{6}*{Protocols} & Yuan \cite{yuan2018blockchain} & 2018 &  Blockchain consensus algorithms \\
    
	\cline{3-5}
	
 \multirow{18}*{\textbf{Essentials}}
    &  {}&\multirow{1}*{Wang \cite{wang2018survey}} & 2018 & Consensus mechanisms and mining management in blockchains\\
    
	\cline{3-5}
    &  {}&\multirow{1}*{Garay \cite{garay2018sok}} & \multirow{1}*{2018} & Consensus Taxonomy in Blockchain Era \\
  
	\cline{3-5}
    {}&&\multirow{1}*{Nguyen \cite{nguyen2018survey}} & \multirow{1}*{2018} & Consensus Algorithms Used in Blockchains \\
  
	\cline{3-5}
    {}&&\multirow{1}*{Wang \cite{wang2019survey}} & \multirow{1}*{2019} &  Consensus  and mining strategy in blockchain networks\\
    
	\cline{3-5}
    {}&&\multirow{1}*{Bano \cite{bano2017consensus}} & \multirow{1}*{2019} &  Consensus in the Age of Blockchains\\

	\cline{3-5}
    {}&&\multirow{1}*{Xiao \cite{xiao2020survey}}& 2020 & Distributed Consensus Protocols for Blockchain Networks\\

	\cline{2-5}
	
	&\multirow{3}*{Smart Contract} &
	\multirow{1}*{Atzei \cite{atzei2016survey}} & 2016 & Attacks on Ethereum smart contracts\\
  
	\cline{3-5}
    {}&&\multirow{1}*{Dwived \cite{dwivedi2019blockchain}} & 2019 & Blockchain-Based Smart-Contract Languages\\	
    
	\cline{3-5}
    {}&&\multirow{1}*{Zheng \cite{zheng2020overview}} & 2020 & Challenges, Advances and Platforms of Smart Contracts\\
  
	\cline{2-5}


    &\multirow{2}*{Sharding} & \multirow{1}*{Wang \cite{wang2019sok}} & 2019 & 
  Sharding on blockchains \\

	\cline{3-5}
	{}&& \multirow{1}*{Yu \cite{Yu2020AccessSharding}} & 2020 & 
  Sharding in Blockchains \\
	
	\cline{2-5}
	
	&\multirow{2}*{Scalability} &  
    \multirow{1}*{Pan \cite{chen2018research}} & 2018 &  Scalability of blockchain technology\\
    
	\cline{3-5}
	{}&&\multirow{1}*{Zhou \cite{zhou2020solutions} }  & 2020 & Solutions to Scalability of Blockchain\\

	\cline{2-5}


    &\multirow{2}*{Cross-chain} &  \multirow{1}*{Zamyatin  \cite{zamyatin2019sok} } & \multirow{1}*{2019} & \multirow{1}*{Cross-ledger Communications}\\
    
    \cline{3-5}
	{}&&\multirow{1}*{\modified{Belchior \cite{belchior2020survey}} }  & \modified{2020} & \modified{Interoperability solutions and problems}\\

	\cline{2-5}
    
 %
 & \multirow{5}*{Security} &  \multirow{1}*{Taylor  \cite{taylor2019systematic} } & 2019 & Blockchain cyber security\\

	\cline{3-5}
	&\multirow{5}*{\& Privacy} &\multirow{1}*{Dasgupta \cite{dasgupta2019survey}}  & 2019 & Security perspective of Blockchain\\

	\cline{3-5}

    {}&&\multirow{1}*{Ma \cite{ma2019survey}} & \multirow{1}*{2019} &  Blockchain technology in crowdsourcing services\\
    
	\cline{3-5}

    {}&&\multirow{1}*{Tariq \cite{tariq2019security}} & 2019 &  Security of big data in blockchain-enabled IoT applications\\

	\cline{3-5}

    {}&&\multirow{1}*{Feng \cite{feng2019survey}} & 2019 & Privacy protection in blockchain systems\\
 
	\cline{3-5}
    {}&&\multirow{1}*{Soni \cite{soni2019comprehensive}} & \multirow{1}*{2019} &  Security, privacy and potential applications of Blockchain\\
 
    \cline{3-5}
    {}&&\multirow{1}*{\modified{Xie  \cite{xie2020blockchain}}} & \multirow{1}*{\modified{2020}} &   \modified{Analysis of vulnerabilities, attacks, and other risks of cloud exchange markets}\\
	

    \hline
	\hline


 \textit{Group-2:}
     & \multirow{1}*{Data Analytics} &  \multirow{1}*{Chen  \cite{chen2018blockchain} } & 2018 & Blockchain data analysis\\

	\cline{2-5}
	
 \textbf{Data Mining}
	& \multirow{1}*{Ponzi Scheme} &\multirow{1}*{Bartoletti \cite{bartoletti2020dissecting}} & 2020 & Dissecting Ponzi schemes on Ethereum\\
    
	\cline{3-5}
    
	\hline	
	\hline

 	\textit{Group-3:}
	& \multirow{1}*{Artificial Intelligence}&  \multirow{1}*{Salah  \cite{salah2019blockchain} } & 2019 & Blockchain for Artificial Intelligence \\

	\cline{3-5}
	
	 \textbf{Decision-}
	{}&\multirow{1}*{(AI)}&\multirow{1}*{Zheng  \cite{zheng2020meetAI}} & 2020 &  Blockchain and Artificial Intelligence\\

	\cline{2-5}

	 \textbf{Making}
	&\multirow{1}*{Machine Learning} &  \multirow{1}*{Chen  \cite{chen2018machine} } & 2018 & Privacy and security design when integrating ML and blockchain \\

	\cline{3-5}
	
	 \textbf{Techniques}
	{}& \multirow{1}*{(ML)} &\multirow{1}*{Liu  \cite{liu2020blockchain}} & 2020 & {Blockchain and ML for Comm. and Networking Systems}\\

	\cline{2-5}
	

    & \multirow{1}*{Game Theory} &  \multirow{1}*{Liu  \cite{liu2019survey} } & 2019 & Game theories on blockchain\\

	\hline
	\hline
	
 	\textit{Group-4:}
	&\multirow{1}*{Cloud Computing} 
	&  Park \cite{park2017blockchain} & 2017 &  {Blockchain security in cloud computing}\\
	
	\cline{2-5}

 	\textbf{New Comm.}
	&\multirow{2}*{Edge Computing} 
	& \multirow{1}*{Xiong  \cite{xiong2018mobile}} & 2018 &  Blockchain meets edge computing\\

	\cline{3-5}
 	\textbf{Networking}
	 {} &&\multirow{1}*{Yang  \cite{yang2019integrated}} & \multirow{1}*{2019} &  Integration of blockchain and edge computing systems\\

	\cline{2-5}
    & \multirow{1}*{5G and Beyond}
     & Nguyen  \cite{nguyen2019blockchain} & 2019 &  Blockchain for 5G and Beyond Networks\\

	\hline

\end{tabular}

\end{table*}

\subsubsection{\textbf{Data Mining and Analytics}}

  The direction of data analytics for blockchains \cite{chen2018blockchain, bartoletti2020dissecting, xie2020blockchain} has not yet received too much attention. The existing survey studies are shown as follows.
  Chen \textit{et al.} \cite{chen2018blockchain} summarized seven typical research issues of data analysis in blockchains, such as entity recognition, privacy identification, network risk parsing, network visualization and portrait, analysis of cryptocurrency market, and etc.
  Recently, Bartoletti \textit{et al.} \cite{bartoletti2020dissecting} reviewed the Ponzi schemes hiding in Ethereum, aiming to discover the scam behavior and analyze their impact. The authors focused on multiple viewpoints such as the identification methods, the impact of Ponzi schemes to the blockchain ecosystem.
  Finally, Xie \textit{et al.} \cite{xie2020blockchain} provided an overview on the security and privacy issues, management of transactions, reputation systems of could exchange, where the blockchain technology is used as a key enabler.

\subsubsection{\textbf{Decision-Making Techniques}}
 
 Blockchains can bring many security advantages for many other fields. On the other hand, blockchain networks also reply on decision-making techniques such as  artificial intelligence (AI) \cite{salah2019blockchain, zheng2020meetAI}, machine learning \cite{chen2018machine, liu2020blockchain}, and game theory \cite{liu2019survey}. This is because the tuning of blockchain network parameters, analysis of user behavior patterns, detection of malicious attacks, identification of market risks, and etc., are playing critical roles for the performance, security, healthy conditions of blockchain systems and blockchain networks. 
 For example,
 Salah  \textit{et al.} \cite{salah2019blockchain} studied how blockchain technologies benefit key problems of AI.
 Zheng \textit{et al.} \cite{zheng2020meetAI} proposed the concept of \textit{blockchain intelligence} and pointed out the opportunities that both these two terms can benefit each other.
 Next, Chen \textit{et al.} \cite{chen2018machine} discussed the privacy-preserving and secure design of machine learning when blockchain techniques are imported.
   Liu  \textit{et al.} \cite{liu2020blockchain} identified the overview, opportunities, and applications when integrating blockchains and machine learning technologies in the context of communications and networking.
    Recently, game theoretical solutions \cite{liu2019survey} have been reviewed when they are applied in blockchain security issues such as malicious attacks and selfish mining, as well as the resource allocation in the management of mining. Both the advantages and disadvantages of game theoretical solutions and models were discussed.

\subsubsection{\textbf{New Communications Networking}}

 First, Park \textit{et al.} \cite{park2017blockchain} discussed how to take the advantages of blockchains in cloud computing with respect to security solutions. Xiong \textit{et al.} \cite{xiong2018mobile} then investigated how to facilitate blockchain applications in mobile IoT and edge computing environments.
 Yang \textit{et al.} \cite{yang2019integrated} identified various perspectives including motivations, frameworks, and functionalities when integrating blockchain with edge computing.
 Nguyen  \textit{et al.} \cite{nguyen2019blockchain} presented a comprehensive survey when blockchain meets 5G networks and beyond. The authors focused on the opportunities that blockchain can bring for 5G technologies, which include cloud computing, mobile edge computing, SDN/NFV, network slicing, D2D communications, 5G services, and 5G IoT applications.

\begin{table*}[t]
\centering
\footnotesize
\caption{Taxonomy of existing blockchain-related surveys (Part 2).}
\label{Table:taxonomyPart2}
\begin{tabular}{|p{0.1\textwidth}|p{0.11\textwidth}|p{0.1\textwidth}|p{0.029\textwidth}|p{0.5\textwidth}|}%
\hline
 \textbf{Group}&\textbf{Category}&\textbf{Ref.}&\textbf{Year} &\textbf{Topic}\\
\hline

  \multirow{12}*{\textit{Group-5:}}
    &\multirow{15}*{IoT, IIoT} & \multirow{1}*{Christidis \cite{christidis2016blockchains}} & 2016 & Blockchains and Smart Contracts for IoT\\
	
	\cline{3-5}
 	 \multirow{12}*{\textbf{IoT \& IIoT}}
     {}&&\multirow{1}*{Ali \cite{ali2018applications}} & 2018 &  Applications of blockchains in IoT\\

	\cline{3-5}
    {}&&\multirow{1}*{Fernandez \cite{fernandez2018review}} & 2018 &  Usage of Blockchain for IoT\\

	\cline{3-5}
    {}&&\multirow{1}*{Kouicem \cite{kouicem2018internet}} & 2018 &  IoT security\\
    
	\cline{3-5}
    {}&&\multirow{1}*{Panarello \cite{panarello2018blockchain}} & 2018 & Integration of  Blockchain and IoT \\

	\cline{3-5}
    {}&&\multirow{1}*{Dai  \cite{dai2019blockchain}} & 2019 &  Blockchain for IoT\\

	\cline{3-5}
	{}&&\multirow{1}*{Wang \cite{wang2019surveyIoT}} & 2019  & Blockchain for IoT\\ 
    
	\cline{3-5}
    {}&&\multirow{1}*{Nguyen \cite{nguyen2019integration}} & 2019 &  Integration of Blockchain and Cloud of Things\\

	\cline{3-5}
    {}&&\multirow{1}*{Restuccia \cite{restuccia2019blockchain}} & 2019 &  Blockchain technology for IoT\\
    
	\cline{3-5}
    {}&&\multirow{1}*{Cao \cite{cao2019internet}} & 2019 &  {Challenges in distributed consensus of IoT}\\
    
	\cline{3-5}
    {}&&\multirow{1}*{Park \cite{park2020greeniot}} & 2020 &  Blockchain Technology for Green IoT\\
    
	\cline{3-5}
    {}&&\multirow{1}*{Lao \cite{lao2020survey}} & \multirow{1}*{2020} & IoT Applications in Blockchain Systems\\
    
	\cline{3-5}
    {}&&\multirow{1}*{Alladi \cite{alladi2019blockchain}} & 2019 &  Blockchain Applications in Industry 4.0 and IIoT\\
    
    	\cline{3-5}
    {}&&\multirow{1}*{Zhang \cite{zhang2019edge}} & 2019 &  {5G Beyond for IIoT based on Edge Intelligence and Blockchain}\\

    \cline{2-5}

    & \multirow{1}*{UAV} & {Alladi  \cite{alladi2020applications}} & 2020 &  Blockchain-based UAV applications\\

	\hline
	\hline
	

	
  \multirow{10}*{\textit{Group-6:}}
	& \multirow{1}*{General} & \multirow{1}*{Lu  \cite{lu2018blockchain}} & 2018 &  Functions, applications and open issues of Blockchain\\

	\cline{3-5}
	 \multirow{10}*{\textbf{Blockchain}}
   & \multirow{1}*{Applications} &\multirow{1}*{Casino \cite{casino2019systematic}} & \multirow{1}*{2019} & Current status, classification and open issues of Blockchain Apps\\
 
	\cline{2-5}
	 \multirow{10}*{\textbf{Applications}}
	&\multirow{2}*{Agriculture} & \multirow{1}*{Bermeo  \cite{bermeo2018blockchain}} & 2018 &  Blockchain technology in agriculture\\
	
	\cline{3-5}
	{} && \multirow{1}*{Ferrag \cite{ferrag2020security}} & \multirow{1}*{2020}  &   Blockchain solutions to Security and Privacy for Green Agriculture\\

	\cline{2-5}
	
	&\multirow{1}*{SDN} & \multirow{1}*{Alharbi \cite{alharbi2020deployment}} & 2020 &  Deployment of Blockchains for Software Defined Networks\\
	
	\cline{2-5}
	
	&\multirow{1}*{Business Apps} & \multirow{1}*{Konst. \cite{konstantinidis2018blockchain}}  & \multirow{1}*{2018} & \multirow{1}*{Blockchain-based business applications} \\

	\cline{2-5}

	& \multirow{1}*{Smart City} & \multirow{1}*{Xie \cite{xie2019survey}} & \multirow{1}*{2019} & Blockchain technology applied in smart cities\\
 
	\cline{2-5}

	& \multirow{2}*{Smart Grids} & \multirow{1}*{Alladi \cite{alladi2019smartgrid}} & 2019 & Blockchain in Use Cases of Smart Grids\\
 
	\cline{3-5}
	{} && \multirow{1}*{Aderibole \cite{aderibole2020blockchain}} & \multirow{1}*{2020}  &  Smart Grids based on Blockchain Technology \\

	\cline{2-5}

	& \multirow{1}*{File Systems} & \multirow{1}*{Huang \cite{huang2020when}} & \multirow{1}*{2020} & \multirow{1}*{Blockchain-based Distributed File Systems, IPFS, Filecoin, etc.}\\
 
	\cline{2-5}

	& \multirow{1}*{Space Industry} & \multirow{1}*{Torky \cite{torky2020blockchain}} & \multirow{1}*{2020} & \multirow{1}*{Blockchain in Space Industry}\\
 
	\cline{2-5}


	& \multirow{1}*{COVID-19} & \multirow{1}*{Nguyen \cite{Nguyen2020covid19}} & \multirow{1}*{2020} & \multirow{1}*{Combat COVID-19 using Blockchain and AI-based Solutions}\\
 
	\hline
	\hline

	
 \multirow{3}*{\textit{Group-7:}}
	& \multirow{4}*{Overview} & \multirow{1}*{Yuan \cite{yuan2016blockchain}} & 2016 &  The state of the art and future trends of Blockchain \\

	\cline{3-5}
    \multirow{3}*{\textbf{General}}
    & \multirow{4}*{\& Outlook} & Zheng \cite{zheng2017overview}  & 2017 & Architecture, Consensus, and Future Trends of Blockchains\\
    
	\cline{3-5}
	\multirow{3}*{\textbf{Overview}}
	{} && \multirow{1}*{Zheng  \cite{zheng2018blockchain}} & 2018 &  Challenges and opportunities of Blockchain\\
   
	\cline{3-5}
	{} && \multirow{1}*{Yuan  \cite{yuan2018blockchainCrypt}} & 2018 &  Blockchain and cryptocurrencies  \\
   
	\cline{3-5}
	{} && \multirow{1}*{Kolb  \cite{kolb2020core}} & 2020 & Core Concepts, Challenges, and Future Directions in Blockchains\\

\hline
	
\end{tabular}

\end{table*}

\subsubsection{\textbf{IoT \& IIoT}}

 The blockchain-based applications for Internet of Things (IoT) \cite{christidis2016blockchains, ali2018applications, fernandez2018review, kouicem2018internet, panarello2018blockchain, dai2019blockchain, wang2019surveyIoT, nguyen2019integration, restuccia2019blockchain, cao2019internet, park2020greeniot, lao2020survey} and Industrial Internet of Things (IIoT) \cite{alladi2019blockchain, zhang2019edge} have received the largest amount of attention from both academia and industry. 
 For example, as a pioneer work in this category, Christidis \textit{et al.} \cite{christidis2016blockchains} provided a survey about how blockchains and smart contracts promote the IoT applications.
  Later on, Nguyen  \textit{et al.} \cite{nguyen2019integration} presented an investigation of the integration between blockchain technologies and cloud of things with in-depth discussion on backgrounds, motivations, concepts and architectures.
  Recently, Park  \textit{et al.}  \cite{park2020greeniot} emphasized on the topic of introducing blockchain technologies to the sustainable ecosystem of green IoT.
  For the IIoT,  Zhang  \textit{et al.} \cite{zhang2019edge} discussed the integration of blockchain and edge intelligence to empower a secure IIoT framework in the context of 5G and beyond.
 In addition, when applying blockchains to the unmanned aerial vehicles (UAV), Alladi  \textit{et al.} \cite{alladi2020applications} reviewed numerous application scenarios covering both commercial and military domains such as network security, surveillance, etc.

\subsubsection{\textbf{Blockchain Applications}}

Blockchains have spawned enormous number of applications in various fields. The research areas covered by the existing surveys on the blockchain-based applications include general applications \cite{lu2018blockchain, casino2019systematic}, agriculture \cite{bermeo2018blockchain, ferrag2020security}, Software-defined Networking (SDN) \cite{alharbi2020deployment}, business applications \cite{konstantinidis2018blockchain}, smart city \cite{xie2019survey}, smart grids \cite{alladi2019smartgrid, aderibole2020blockchain}, distributed file systems \cite{huang2020when}, space industry \cite{torky2020blockchain}, and COVID-19 \cite{Nguyen2020covid19}.
Some of those surveys are reviewed as follows.

Lu \textit{et al.} \cite{lu2018blockchain} performed a literature review on the fundamental features  of blockchain-enabled applications. Through the review, the authors expect to outlook the development routine of blockchain technologies. 
Then, Casino  \textit{et al.} \cite{casino2019systematic} presented a systematic survey of blockchain-enabled applications in the context of multiple sectors and industries. Both the current status and the prospective characteristics of blockchain technologies were identified.
 In more specific directions, Bermeo \textit{et al.}  \cite{bermeo2018blockchain} proposed a review on the research works focusing on applying blockchain technologies to agriculture. Through an overview on the primary studies published between 2016 and 2018, they found some interesting phenomena such as a large part of relevant papers are solving problems of food supply chain, and Asian community researchers are dominating the blockchain-based agriculture studies.
 Later on, Ferrag \textit{et al.}  \cite{ferrag2020security} concentrated on the security and privacy issues of green IoT-based agriculture. They also investigated how would blockchain solutions and consensus algorithms be adapted to green IoT-based agriculture.
Alharbi  \cite{alharbi2020deployment} then described how blockchain technologies can be integrated into SDN architecture to provide security, confidentiality, and integrity.
Konstantinidis \textit{et al.} \cite{konstantinidis2018blockchain} discussed the various applications of blockchain technology on the business sectors.

 Xie \textit{et al.} \cite{xie2019survey} provided a literature review on the smart city services involving blockchain technologies, such as smart citizen, smart healthcare, smart transportation, management of supply chain, etc.
 Then, based on the blockchain technology, the two surveys \cite{alladi2019smartgrid, aderibole2020blockchain} discussed the conceptual model, different use cases, energy trading processes, efficient power generation and distribution strategies, system maintenance and diagnosis for grid facilities, and security and privacy preserving of smart grid domains.
 Huang  \textit{et al.}  \cite{huang2020when} reviewed the integration of blockchain-based solutions and the distributed file systems. Taking the Inter-Planetary File System (IPFS) and Swarm as two representative distributed file systems, the authors introduced the principle and structure, as well as the state-of-the-art studies of blockchain-empowered distributed file systems and their utilization scenarios.
 %
 Next, Torky \textit{et al.}  \cite{torky2020blockchain} conducted a systematic discussion on the conceptual exploration to adopt the blockchain technology in space industry. A blockchain-based satellite network, namely SpaceChain, has been initially implemented as a case study of the proposed blockchain-empowered satellite system.
 As a most timely survey regarding combating the coronavirus (COVID-19), Nguyen \textit{et al.} \cite{Nguyen2020covid19} presented a comprehensive review on the integrating blockchain and AI technologies while fighting the coronavirus crisis. The roles of blockchain during tackling the pandemic vary in a wide range of applications, such as tracking of population, privacy preserving of citizens, supply chain management, and other tracking services.

\subsubsection{\textbf{General Overview \& Outlook}}
 
 The final group of survey articles \cite{yuan2016blockchain, zheng2017overview, zheng2018blockchain, yuan2018blockchainCrypt, kolb2020core} overviewed the basic concepts of blockchains and cryptocurrencies, the fundamental research challenges, and general issues such as consensus algorithms, solutions to scalability and security, privacy preserving issues, and etc. Finally, The authors outlooked further potential technical challenges and open issues for shedding light on future studies of blockchain technologies.
 
\textbf{Summary of Survey-Article Review}: Through the brief review of the state-of-the-art surveys, we have found that the blockchain technologies have been adaptively integrated into a growing range of application sectors. The blockchain theory and technology will bring substantial innovations, incentives, and a great number of application scenarios in diverse fields.
 Based on the analysis of those survey articles, we believe that there will be more survey articles published in the near future, very likely in the areas of sharding techniques, scalability, interoperability, smart contracts, big data, AI technologies, 5G and Beyond,  edge computing, cloud computing, and many other fields.


 \subsection{Motivation of This Survey}
 
 {\color{black}
 
  Via the overview, shown in Table \ref{Table:taxonomy}, Table \ref{Table:taxonomyPart2}, Fig. \ref{fig:category} and Fig. \ref{fig:classification}, of the existing blockchain-related surveys, we have found that a survey of the state-of-the-art theories, modelings and useful tools that can i) improve the performance of blockchains, and ii) help better understand blockchains, is still missing.
 In particular, the following directions need in-depth investigations.


 \subsubsection{Theories to Improving the Performance of Blockchains}
 
 The performance of blockchains includes a number of metrics such as throughput, latency, storage efficiency, reliability, scalability, interoperability, and etc. Many theories can be devoted to improving the performance metrics of blockchains. For example, the following perspectives are worthy paying more efforts.

 \textbf{Scalability Solutions}. Although blockchain is viewed as a distributed and public database of transactions and has become a platform for decentralized applications, blockchains still face the scalability problem. For example, the system throughput is not scalable with the increasing size of a blockchain network. Thus, the scalability solutions of blockchain are still required further studying. The promising solutions to improve the scalability of blockchains include sharding-based and multiple-chain \& cross-chain techniques.

 \textbf{New Protocols and Infrastructures}.  
  Several classic consensus protocols, such as practical byzantine-fault tolerant (PBFT) protocol and proof-of-work (PoW) protocol, have been widely adopted by popular blockchain systems. However, those classic protocols cannot meet all consensus requirements existing in emerging new blockchains. Thus, it is necessary to review new protocols and infrastructures proposed to serve new scenarios of blockchain-based applications.

  \subsubsection{Modelings and Techniques for Better Understanding Blockchains}
 
 The existing studies on better understanding blockchains that have been reviewed by other surveys mainly focus on the security and privacy issues, and the analysis of cryptocurrency market, for example, the identification of Ponzi schemes and other scam behaviors.
 In our opinion, to better understand blockchains, the following wider range of topics should be also emphasized on.

 \textbf{Graph-based Theories}.  Excepting the classic graph \modified{knowledge} that have applied to blockchains, such as the Merkel tree and directed acyclic graph (DAG) techniques, the general graph based analytical techniques are powerful approaches to find insights behind the transactions, smart contracts, and the network structure of blockchains. 
 
 \textbf{Stochastic Modelings, Queueing Theories and Analytical Models}. 
 Several phases of blockchain networks can be described using the stochastic modelings, queueing theories, and analytical models. Based on these theoretical models, researchers can conduct the property analysis of blockchain network, stability analysis, deriving failure probability, modeling of mining procedure,  estimating blockchain confirming time, exploring the synchronization process of Bitcoin network and other working principles of blockchains, and understanding how blockchains respond to difference network conditions even malicious attacks.

 \textbf{Data Analytics for Cryptocurrency Blockchains}. 
 Security issues of cryptocurrency blockchains and their markets are attracting more and more attention. Although several surveys \cite{chen2018blockchain, bartoletti2020dissecting, xie2020blockchain} have already studied the Ponzi schemes in Ethereum and other general security and privacy issues of blockchain systems, their surveys mainly emphasized on the identification approaches and the impacts to blockchain systems.
 In contrast, in our survey, we review the latest studies by exploiting the data analytics techniques to detect the market risks in cryptocurrency ecosystems, where the risks not only include Ponzi schemes, but also take into account the cryptojacking, market manipulation mining, and money-laundering activities. Furthermore, we also review a few studies utilizing data science and stochastic modelings to produce a portrait of cryptoecomonic systems.

  \subsubsection{Useful Measurements, Datasets and Experiment Tools for Blockchains}
 In the aforementioned 66 surveys, we find that there is still no a single article focusing on the performance measurements, datasets and experiment tools for blockchains. Instead, our survey in this article particularly reviews:  i) performance measurements with respect to throughput, end-to-end confirmation delays of transactions, forking rate, resource utilization, scalability, etc.; and ii) useful evaluation tools and datasets dedicated to blockchain experiments.


 In a summary, by this article, we would like to fill the gap by emphasizing on the cutting-edge theoretical studies, modelings, and useful tools for blockchains.
 Particularly, we try to include the latest high-quality research outputs that have not been included by other existing survey articles.
 We believe that this survey can shed new light on the further development of blockchains.

\subsection{Contribution of Our Survey}
Our survey presented in this article includes the following contributions.
\begin{itemize}
	\item We conduct a brief classification of existing blockchain surveys to highlight the meaning of our literature review shown in this survey.
	\item We then present a comprehensive investigation on the state-of-the-art theoretical modelings, analytics models, performance measurements, and useful experiment tools for blockchains, blockchain networks, and blockchain systems.
	\item Several promising directions and open issues for future studies are also envisioned finally.
\end{itemize}

    \begin{figure}[t]
    \centering
	\vspace{-3mm}
    \includegraphics[width=0.56\textwidth]{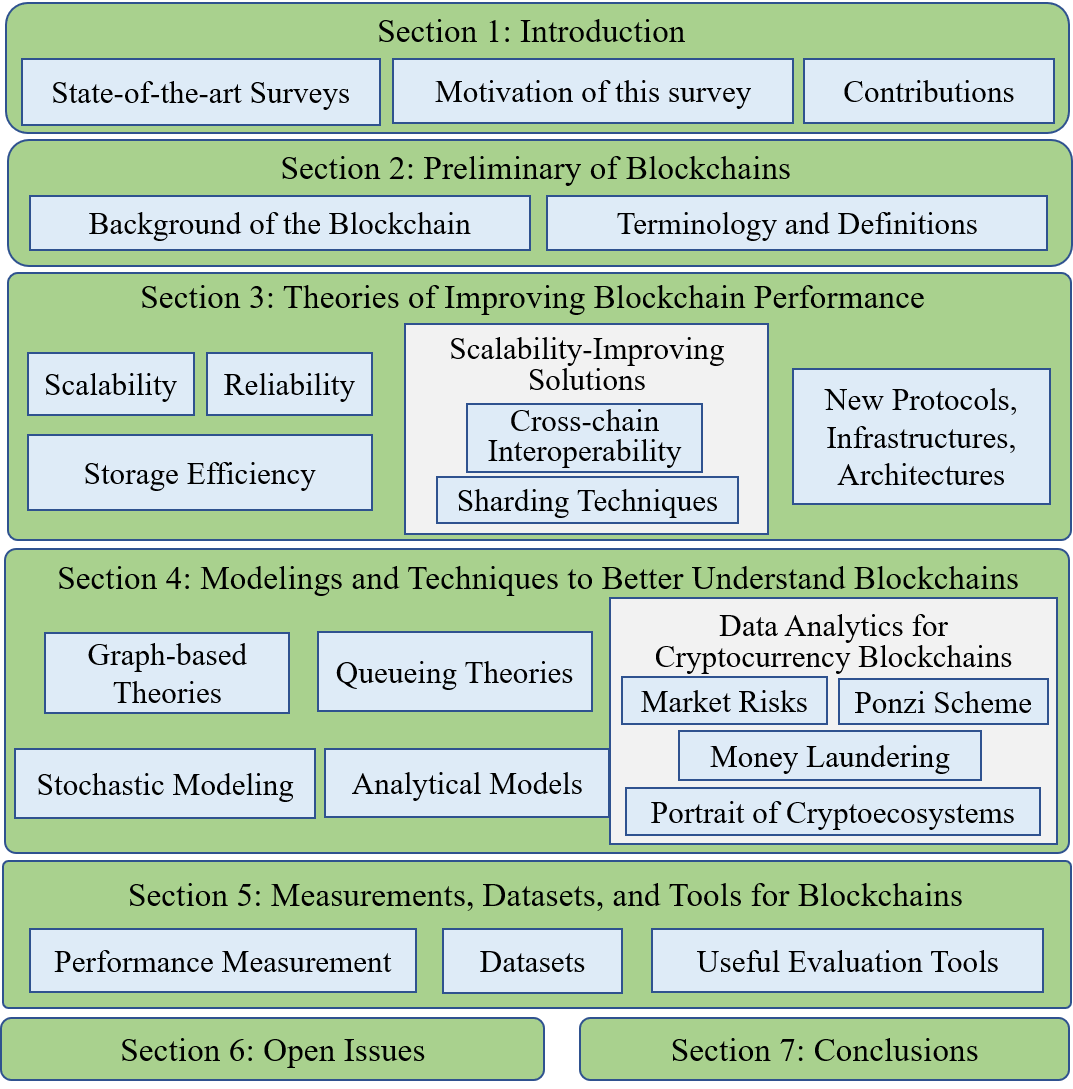}
	\vspace{-2mm}
    \caption{The structure of this article.}
    \label{fig:structure}
    \vspace{-5mm}
    \end{figure}


The structure of this survey is shown in Fig. \ref{fig:structure} and organized as follows.
Section \ref{sec:preliminary} introduces the preliminaries of blockchains.
Section \ref{sec:theoryImproving} summarizes the state-of-the-art theoretical studies that improve the performance of blockchains.
In Section \ref{sec:understand}, we then review various modelings and analytic models that help understand blockchains.
Diverse measurement approaches, datasets, and useful tools for blockchains are overviewed in Section \ref{sec:tools}.
We outlook the open issues in Section \ref{sec:openissue}.
Finally, Section \ref{sec:conclusion} concludes this article. 

}


{\color{black}

\section{Preliminaries of Blockchains}\label{sec:preliminary}

 Blockchain is a promising paradigm for content distribution and distributed consensus over P2P networks. In this section, we present the basic concepts, definitions and terminologies of blockchains appeared in this article. \modified{Due to the frequent use of acronyms in this paper, we will include an acronym table, i.e., Table \ref{Table:acronyms}, in this section.}
 
\begin{table*}[t]
\caption{\modified{Acronym Table}}
\centering
\footnotesize
\begin{tabular}{|p{0.25\textwidth}|p{0.25\textwidth}|}
\hline
\textbf{Acronym} &\textbf{Meaning}\\
    \hline
    AI & Artificial Intelligence\\
    \hline
    BFT & Byzantine Fault Tolerant\\
    \hline
    CA & Contract Account\\
    \hline
    CapsNet & Capsule Network\\
    \hline
    CCG & Contract Creation Graph\\
    \hline
    CIG & Contract Invocation Graph\\
    \hline
    DAG & Directed Acyclic Graph\\
    \hline
    DApp & Distributed Application\\
    \hline
    EHG & Extreme High Graph\\
    \hline
    ELG & Extreme Low Graph\\
    \hline
    EOA & External Owned Account\\
    \hline
    ETG & Extreme Transaction Graph\\
    \hline
    EVM & Ethereum Virtual Machine\\
    \hline
    IIoT & Industrial Internet of Things\\
    \hline
    IoT & Internet of Things\\
    \hline
    IPFS & Inter-Planetary File System\\
    \hline
    L2S & Latency-to-Shard\\
    \hline
    MFG & Money Flow Graph\\
    \hline
    ML & Machine Learning\\
    \hline
    MMR & Monitor Multiplexing Reading\\
    \hline
    NMG & Normal Graph\\
    \hline
    PBFT & Practical Byzantine-fault Tolerant\\
    \hline
    PoS & Proof of Stake\\
    \hline
    PoT & Proof-of-Trust\\
    \hline
    PoW & Proof of Work\\
    \hline
    RDMA & Remote Direct Memory Access\\
    \hline
    SDN & Software-defined Networking\\
    \hline
    SPV & Simple Payment Verification\\
    \hline
    T2S & Transaction-to-Shard\\
    \hline
    TX & Transaction\\
    \hline
    UAV & Unmanned Aerial Vehicles \\
    \hline
    UTXO & Unspent Transaction Output\\

    \hline

	
	  
	
\end{tabular}
\label{Table:acronyms}
\end{table*}

 \subsection{Prime Blockchain Platforms}

 \subsubsection{Bitcoin}
 
  Bitcoin is viewed as the blockchain system that executes the first cryptocurrency. It builds upon two major techniques, i.e., \textit{Nakamoto Consensus} and \textit{UTXO Model}, which are introduced as follows.
  
  \textbf{Nakamoto Consensus}. To achieve an agreement of blocks, Bitcoin adopts the Nakamoto Consensus, in which miners generate new blocks by solving a puzzle. In such a puzzle-solving process, also referred to as mining, miners need to calculate a nonce value that fits the required difficulty level  \cite{nakamoto2008bitcoin}. Through changing the difficulty, Bitcoin system can maintain a stable rate of block-generation, which is about one block per 10 minutes.
  When a miner generates a new block, it broadcasts this message to all the other miners in the network. If others receive this new block, they add this block to their local chain. If all of the other miners receive this new block timely, the length of the main chain increases by one. However, because of the network delays, not always all the other miners can receive a new block in time. When a miner generates a block before it receives the previous one, a fork yields. Bitcoin addresses this issue by following the rule of longest chain.
  
  \textbf{UTXO Model}. The Unspent Transaction Output (UTXO) model is adopted by cryptocurrencies like Bitcoin, and other popular blockchain systems \cite{luu2016secure, kokoris2018omniledger}. A UTXO is a set of digital money, each represents a chain of ownership between the owners and the receivers based on the cryptography technologies.
  In a blockchain, the overall UTXOs form a set, in which each element denotes the unspent output of a transaction, and can be used as an input for a future transaction. A client may own multiple UTXOs, and the total coin of this client is calculated by summing up all associated UTXOs.
   Using this model, blockchains can prevent the double-spend \cite{moroz2020doublespend} attacks efficiently.

 \subsubsection{Ethereum}  
 Ethereum \cite{wood2014ethereum} is an open-source blockchain platform enabling the function of smart contract.
 As the token in Ethereum, \textit{Ether} is rewarded to the miners who conducted computation to secure the consensus of the blockchain.
 Ethereum executes on decentralized Ethereum Virtual Machines (EVMs), in which scripts are running on a network consisting of public Ethereum nodes. Comparing with Bitcoin, the EVM's instruction set is believed Turing-complete. Ethereum also introduces an internal pricing mechanism, called \textit{gas}. A unit of gas measures the amount of computational effort needed to execute operations in a transaction. Thus, gas mechanism is useful to restrain the spam in smart contracts.
 Ethereum 2.0 is an upgraded version based on the original Ethereum. The upgrades include a transition from PoW to Proof-of-Stake (PoS), and a throughput-improving based on  sharding technologies.
 \modified{The comparison between Bitcoin \& Ethereum is summarized in Table \ref{Table:compareBandE}.}
 
 \modified{\textbf{Account/Balance Model}. Unlike Bitcoin where states are composed by UTXOs, Ethereum adopts a more common and straightforward model that is used by banks, the Account/Balance Model. In every account, an incrementing counter of transaction execution, nonce, is implemented to prevent double spending attacks, which serves as a complement for the model's simple structure. There are basically 2 types of accounts, \textit{external owned accounts} (EOAs) and \textit{contract accounts} (CAs), each controlled by private keys and contract codes, respectively.}
\begin{table*}[t]
\caption{\modified{Comparison between Bitcoin \& Ethereum}}
\centering
\footnotesize
\begin{tabular}{|p{0.1\textwidth}|p{0.13\textwidth}|p{0.17\textwidth}|p{0.1\textwidth}|}
\hline
\textbf{} &\textbf{State Model} &\textbf{Consensus Protocols} &\textbf{Throughput}\\
\hline

	%
	\textbf{Bitcoin}
	&  UTXO    & PoW & 3 to 7 TPS\cite{ethTPS}\\
	
	\hline
    \textbf{Ethereum1.0}
	& Account/Balance & PoW  & 7 to 15 TPS\cite{ethTPS}\\
	  
	\hline
	\textbf{Ethereum2.0}
	& Account/Balance  & PoS Sharding  & Unknown\\
	\hline
	
\end{tabular}
\label{Table:compareBandE}
\end{table*}
\modified{\subsubsection{Hyperledger Fabric}
Hyperledger Fabric \cite{hyperledgerfabric} is a popular permissioned blockchain platform for industrial use. In industry, goals are quite different from cryptocurrency systems. Greater significance is attached to lower maintenance cost, higher throughput performance and permission control. For a node in a permissioned setting, other nodes, though untrusted, the identities are known. With different levels of trust among users, different consensus protocols can be customized for fault tolerant.
} 
 
 \subsubsection{EOSIO}
 
 EOSIO \cite{EOSIO} is another popular blockchain platform released by a company \textit{block.one}  on 2018. Different from Bitcoin and Ethereum, the smart contracts of EOSIO don't need to pay transaction fees. Its throughput is claimed to reach millions of transactions per second. Furthermore, EOSIO also enables low block-confirmatoin latency, low-overhead BFT finality, and etc. These excellent features has attracted a large-number of users and developers to quickly and easily deploy decentralized applications in a governed blockchain. For example, in total 89,800,000 EOSIO blocks have been generated in less than one and a half years since its first launching.


 \subsection{Consensus Mechanism}
 
 The consensus mechanism in blockchains is for fault-tolerant to achieve an agreement on the same state of the blockchain network, such as a single state of all transactions in a cryptocurrency blockchain.
 Popular proof-based consensus protocols include PoW and PoS.
 In PoW, miners compete with each other to solve a puzzle that is difficult to produce a result but easy to verify the result by others. Once a miner yields a required nonce value through a huge number of attempts, it gets paid a certain cryptocurrencies for creating a new block.
 In contrast, PoS doesn't have miners. Instead, the new block is forged by \textit{validators} selected randomly within a committee. The probability to be chosen as a validator is linearly related to the size of its stake.
 PoW and PoS are both adopted as consensus protocols for the security of cryptocurrencies. The former is based on the CPU power, and the latter on the coin age. Therefore, PoS is with lower energy-cost and less likely to be attacked by the 51\% attack.


 \subsection{Scalability of Blockchains}

 Blockchain as a distributed and public database of transactions has become a platform for decentralized applications. Despite its increasing popularity, blockchain technology faces the scalability problem: throughput does not scale with the increasing network size. Thus, scalable blockchain protocols that can solve the scalability issues are still in an urgent need. Many different directions, such as \textit{Off-chain}, \textit{DAG}, and \textit{Sharding} techniques, have been exploited to address the scalability of blockchains.
Here, we present several representative terms related to scalability.

 \subsubsection{Off-chain Techniques}
Contrary to the on-chain transactions that are dealt with on the blockchain and visible to all nodes of the blockchain network, the off-chain transactions are processed outside the blockchain through a third-party guarantor who endorses the correctness of the transaction.  
 The on-chain transactions incur longer latencies since the confirmation of an on-chain transaction has to take different steps. In contrast, the off-chain techniques can instantly execute the off-chain transactions because those transactions don't need to wait on the queue as on an on-chain network.

 \subsubsection{DAG}
 Mathematically, a DAG is a finite directed graph where no directed cycles exist. In the context of blockchain, DAG is viewed as a revolutionized technology that can upgrade blockchain to a new generation. This is because DAG is blockless, and all transactions link to multiple other transactions following a topological order on a DAG network. Thus, data can move directly between network participants. This results in a faster, cheaper and more scalable solution for blockchains.
 In fact, the bottleneck of blockchains mainly relies on the structure of blocks. Thus, probably the blockless DAG could be a promising solution to improve the scalability of blockchains substantially.

 \subsubsection{Sharding Technique}

 The consensus protocol of Bitcoin, i.e., Nakamoto Consensus, has significant drawbacks on the performance of transaction throughput and network scalability. To address these issues, \textit{sharding} technique is one of the outstanding approaches, which improves the throughput and scalability by partitioning the blockchain network into several small shards such that each can process a bunch of unconfirmed transactions in parallel to generate medium blocks. Such medium blocks are then merged together in a final block.  
 Basically, sharding technique includes \textit{Network Sharding}, \textit{Transaction Sharding} and \textit{State Sharding}.

 \subsubsection{Cross-Shard Transactions}
  
  One shortcoming of sharding technique is that the malicious network nodes residing in the same shard may collude with each other, resulting in security issues. Therefore, the sharding-based protocols exploits \textit{reshuffling} strategy to address such security threats. However, reshuffling brings the \textit{cross-shard} data migration. Thus, how to efficiently handle the cross-shard transactions becomes an emerging topic in the context of sharding blockchain.

}


\section{Theories to Improving the Performance of Blockchains}\label{sec:theoryImproving}

 \subsection{ {\color{black} Latest Theories to Improving Blockchain Performance}}
 \modified{Summary of this subsection is included in Table \ref{Table:theories}.}


 \subsubsection{ {\color{black} Throughput \& Latency}}

    {\color{black} 
    
 Aiming to reduce the confirmation latency of transactions to milliseconds,  Hari \textit{et al.} \cite{hari2019accel} proposed a high-throughput, low-latency, deterministic confirmation mechanism called ACCEL for accelerating Bitcoin's block confirmation. The key findings of this paper includes how to identify the singular blocks, and how to use singular blocks to reduce the confirmation delay. Once the confirmation delay is reduced, the throughput increases accordingly.

 Two obstacles have hindered the scalability of the cryptocurrency systems. The first one is the low throughput, and the other one is the requirement for every node to duplicate the communication, storage, and state representation of the entire blockchain network. Wang \textit{et al.} \cite{wang2019monoxide}  studied how to solve the above obstacles. Without weakening decentralization and security, the proposed Monoxide technique offers a linear scale-out ability by partitioning the workload. And they preserved the simplicity of the blockchain system and amplified its capacity. The authors also proposed a novel \textit{Chu-ko-nu} mining mechanism, which ensures the cross-zone atomicity, efficiency and security of the blockchain system with thousands of independent zones. Then, the authors have conducted experiments to evaluate the scalability performance of the proposed Monoxide with respect to TPS, the overheads of  cross-zone transactions, the confirmation latency of transactions, etc.

    To bitcoin, low \textit{throughput} and long \textit{transaction confirmation latency} are two critical bottleneck metrics. To overcome these two bottlenecks, Yang \textit{et al.}  \cite{yang2019prism} designed a new blockchain protocol called Prism, which achieves a scalable throughput as high as 70,000 transactions per second, while ensuring a full security of bitcoin. 
    The project of Prism is open-sourced in Github. The instances of Prism can be flexibly deployed on commercial cloud platform such as AWS.
    However, the authors also admitted that although the proposed Prism has a high throughput, its confirming latency still maintains as large as 10 seconds since there is only a single \textit{voter chain} in Prism. A promising solution is to introduce a large number of such voter chains, each of which is not necessarily secure. Even though every voter chain is under attacking with a probability as high as 30\%, the successful rate of attacking a half number of all voter chains is still theoretically very low. Thus, the authors believed that using multiple voter chains would be a good solution to reducing the confirmation latency while not sacrificing system security.

 Considering that Ethereum simply allocates transactions to shards according to their account addresses rather than relying on the workload or the complexity of transactions, the resource consumption of transactions in each shard is unbalanced. In consequence, the network transaction throughput is affected and becomes low. To solve this problem, Woo \textit{et al.}  \cite{woo2020garet} proposed a heuristic algorithm named GARET, which is a gas consumption-aware relocation mechanism for improving throughput in sharding-based Ethereum environments. In particular, the proposed GARET can relocate transaction workloads of each shard according to the gas consumption. The experiment results show that GARET achieves a higher transactions throughput and a lower transaction latency compared with existing techniques.

    }

\begin{table*}[t]
\caption{Latest Theories of Improving the Performance of Blockchains.}
\centering
\footnotesize
\begin{tabular}{|p{0.08\textwidth}|p{0.03\textwidth}|p{0.12\textwidth}|p{0.23\textwidth}|p{0.35\textwidth}|}
\hline
\textbf{Emphasis} &\textbf{Ref.} &\textbf{Recognition} &\textbf{{\color{blue}Challenge}}& \textbf{Methodology}\\
\hline


	%
	\multirow{11}*{Throughput} 
	&     \cite{hari2019accel}   & ACCEL: Reduce the confirmation delay of blocks & {\color{blue}Most of the blockchain applications desire fast confirmation of their transactions} & Authors proposed a high-throughput, low-latency, deterministic confirmation mechanism, aiming to accelerate Bitcoin's block confirmation. \\

	\cline{2-5}
	\multirow{6}*{\& Latency} &    \cite{wang2019monoxide}  & Monoxide & {\color{blue}Scalability issues, and efficient processing of cross-shard transactions} &  The proposed Monoxide offers a linear scale-out by partitioning workloads. Particularly, \textit{Chu-ko-nu} mining mechanism enables the cross-zone atomicity, efficiency and security of the system. \\
	
	\cline{2-5}
	{ } &   \cite{yang2019prism}  & Prism  & {\color{blue}Low transaction throughput and large transaction confirmation of bitcoin}  & Authors proposed a new blockchain protocol, i.e., Prism, aiming to  achieve a scalable throughput with a full security of bitcoin. \\

	\cline{2-5}
	&   \cite{woo2020garet}  & GARET  & {\color{blue}How to place transactions to shards considering the complexity of  transactions or the workload generated by transactions} & Authors proposed a gas consumption-aware relocation mechanism for improving throughput in sharding-based Ethereum. \\

	\hline
	
	\multirow{8}*{Storage} 
	&   \cite{perard2018erasure}& Erasure code-based & {\color{blue}How to reduce the storage consumption of blockchains} & Authors proposed a new type of low-storage blockchain nodes using erasure code theory to reduce the storage space of blockchains. \\

	\cline{2-5}
	\multirow{4}*{Efficiency} 
	&   \cite{dai2019jidar} & Jidar: Data-Reduction Strategy & {\color{blue}How to reduce the data consumption of bitcoin's blocks} &  Authors proposed a data reduction strategy for Bitcoin namely Jidar, in which each node only has to store the transactions of interest and the related Merkle branches from the complete blocks. \\
	 
	\cline{2-5}
	&   \cite{xu2020segmentblockchain}  & \textit{Segment blockchain}  & {\color{blue} To reduce the storage of blockchain systems while maintaining the decentralization without sacrificing security} & Authors proposed a data-reduced storage mechanism named \textit{segment blockchain} such that each node only has to store a segment of the blockchain. \\

	\hline
	\multirow{5}*{Reliability} 
	&   \cite{weber2017availability}  & Availability of blockchains & {\color{blue} The availability of read and write on blockchains is uneven} &  Authors studied the availability for blockchain-based systems, where the read and write availability is conflict to each other. \\
	
	\cline{2-5}
	\multirow{1}*{Analysis} 
	&   \cite{zheng2019selecting}  & Reliability prediction  & {\color{blue}The reliability of blockchain peers is unknown} &  Authors proposed H-BRP to predict the reliability of blockchain peers by extracting their reliability parameters. \\
	 
	\hline
	
\end{tabular}
\label{Table:theories}
\end{table*}
\begin{table*}[t]
\caption{Latest Scalability Solutions to Improving the Performance of Blockchains.}
\centering
\footnotesize
\begin{tabular}{|p{0.12\textwidth}|p{0.03\textwidth}|p{0.12\textwidth}|p{0.62\textwidth}|}%
\hline
\textbf{Emphasis} &\textbf{Ref.} &\textbf{Recognition} &\textbf{Methodology}\\
\hline

	%
	\multirow{24}*{Solutions to}
	&   \cite{luu2016secure}   & Elastico & Authors proposed a new distributed agreement protocol for the permission-less blockchains, called Elastico, which is viewed as the first secure candidate for a sharding protocol towards the open public blockchains.\\
	
	\cline{2-4}
	\multirow{20}*{Sharding} &    \cite{wang2019monoxide}  & Monoxide & \modified{The proposed Monoxide enables the system to handle transactions through a number of independent zones. This scheme is essentially following the principle of sharding mechanism.} \\

	\cline{2-4}
	\multirow{18}*{blockchains}
	&   \cite{zamani2018rapidchain}   & Rapidchain & Authors proposed a new sharding-based protocol for public blockchains that achieves non-linearly increase of intra-committee communications with the number of committee \modified{members}.\\
	
	\cline{2-4}
	{ } &    \cite{amiri2019sharper}  & SharPer & Authors proposed a permissioned blockchain system named \textit{SharPer}, which adopts sharding techniques to improve scalability of cross-shard transactions.\\
	
	\cline{2-4}
	{ } &    \cite{kim2019gas} & D-GAS  & Authors proposed a dynamic load balancing mechanism for Ethereum shards, i.e., D-GAS. It reallocates Tx accounts by their gas consumption on each shard.\\
	
	\cline{2-4}
	{ } &   \cite{Wang2019sharding} & NRSS  & Authors proposed a node-rating based new Sharding scheme, i.e., NRSS, for blockchains, aiming to improve the throughput of committees.\\
	
	\cline{2-4}
	{ } &   \cite{nguyen2019optchain} & OptChain  & Authors proposed a new sharding paradigm, called OptChain, mainly used for optimizing the placement of transactions into shards.\\

	\cline{2-4}
	{ } &   \cite{dang2019towards}  & Sharding-based scaling system & Authors proposed an efficient shard-formation protocol that assigns nodes into shards securely, and a distributed transaction protocol that can guard against malicious Byzantine fault \modified{coordinators}.\\
	
	\cline{2-4}
	{ } &   \cite{chen2019sschain} & SSChain  & Authors proposed a non-reshuffling structure called SSChain, which supports both transaction sharding and state sharding while eliminating huge data-migration across shards.\\

	\cline{2-4}
	{ } &   \cite{niu2019eunomia}  & Eumonia  & Authors proposed Eumonia, which is a permissionless parallel-chain protocol for realizing a global ordering of blocks.\\
	 
	\cline{2-4}
	{ } &   \cite{rajab2020feasibility}  & Vulnerability of Sybil attacks  & Authors systematically analyzed the vulnerability of Sybil attacks in protocol Elastico.\\

	\cline{2-4}
	{ } &    \cite{xu2020n}  & n/2 BFT Sharding approach  & Authors proposed a new blockchain sharding approach that can tolerate up to 1/2 of the  Byzantine nodes within a shard.\\
	
	\cline{2-4}
	{ } &    \cite{zhang2020cycledger}  & CycLedger  & Authors proposed a protocol CycLedger to pave a way towards scalability, security and incentive for sharding blockchains.\\

	\hline
	\multirow{10}*{Interoperability }
	&    \cite{jin2018towards} & Interoperability architecture & Authors proposed a novel interoperability architecture that supports the cross-chain cooperations among multiple blockchains, and a novel Monitor Multiplexing Reading (MMR) method for the passive cross-chain communications.\\

	\cline{2-4}
	\multirow{6}*{of multiple-chain}
	&    \cite{liu2019hyperservice}  & HyperService  & Authors proposed a programming platform that provides interoperability and programmability over multiple heterogeneous blockchains.\\
	
	\cline{2-4}
	\multirow{4}*{systems}
	&    \cite{fynn2020smart}  & Protocol \textit{Move}  & Authors proposed a programming model for smart-contract developers to create DApps that can interoperate and scale in a multiple-chain \modified{environment}.\\
	 
	\cline{2-4}
	{ } &    \cite{tian2020enabling}  & Cross-cryptocurrency TX protocol  & Authors proposed a decentralized cryptocurrency exchange protocol enabling cross-cryptocurrency transactions based on smart contracts deployed on Ethereum.\\
	 
	\cline{2-4}
	{ } &    \cite{zamyatin2019sok}  & Cross-chain comm.  & Authors conducted a systematic classification of cross-chain communication protocols.\\
	 
	\hline

\end{tabular}
\label{Table:scalability}
\end{table*}
 
 
 \subsubsection{{\color{black} Storage Efficiency}}
 
 {\color{black}
 
 The transactions generated at real-time make the size of blockchains keep growing. For example, the storage efficiency of original-version Bitcoin has received much criticism since it requires to store the full transaction history in each Bitcoin peer. Although some revised protocols advocate that only the full-size nodes store the entire copy of whole ledger, the transactions still consume a large storage space in those full-size nodes. To alleviate this problem, several pioneer studies proposed storage-efficient solutions for blockchain networks.
 For example,
 By exploiting the erasure code-based approach, Perard  \textit{et al.} \cite{perard2018erasure} proposed a low-storage blockchain mechanism, aiming to achieve a low requirement of storage for blockchains. The new low-storage nodes only have to store the linearly encoded fragments of each block. The original blockchain data can be easily recovered by retrieving fragments from other nodes under the erasure-code framework. Thus, this type of blockchain nodes allows blockchain clients to reduce the storage capacity. The authors also tested their system on the low-configuration Raspberry Pi to show the effectiveness, which demonstrates the possibility towards running blockchains on IoT devices.
 
 Then, Dai  \textit{et al.} \cite{dai2019jidar} proposed Jidar, which is a data reduction strategy for Bitcoin. In Jidar, each node only has to store the transactions of interest and the related Merkle branches from the complete blocks. All nodes verify transactions collaboratively by a query mechanism. This approach seems very promising to the storage efficiency of Bitcoin. Their experiments show that the proposed Jidar can  reduce the storage overhead of each peer \modified{to} about 1\% comparing with the original Bitcoin. 
 
 Under the similar idea, Xu \textit{et al.} \cite{xu2020segmentblockchain} reduced the storage of blockchains using a \textit{segment blockchain} mechanism, in which each node only needs to store a piece of blockchain segment. The authors also proved that the proposed mechanism endures a failure probability $(\phi/n)^m$ if an adversary party commits a collusion with less than a number $\phi$ of nodes and each segment is stored by a number $m$ of nodes. This theoretical result is useful for the storage design of blockchains when developing a particular segment mechanism towards data-heavy distributed applications.
 
 }

 \subsubsection{{\color{black} Reliability of Blockchains}}

 {\color{black}
 \modified{As a decentralized mechanism for data protection, the reliability of blockchains plays an important role in data falsification. The following works studied the fundamental supporting mechanisms to achieve data falsification prevention.}
 The availability of blockchains is a key factor for blockchain-based distributed applications (DApps). However, such availability guarantees of blockchain systems are unknown. To this end, Weber \textit{et al.} \cite{weber2017availability} studied the availability limitations of two popular blockchains, i.e., Bitcoin and Ethereum. The authors found that the availability of reading and writing operations are conflict to each other. Through measuring and analyzing the transactions of Ethereum, they observed that the DApps could be stuck in an uncertain state while transactions are pending in a blockchain system. This observation suggests that maybe blockchains should support some built-in transaction-abort options for DApps. The authors finally presented techniques that can alleviate the availability limitations of Ethereum and Bitcoin blockchains.

 In public blockchains, the system clients join the blockchain network basically through a third-party peer. Thus, the reliability of the selected blockchain peer is critical to the security of clients in terms of both resource-efficiency and monetary issues. To enable clients evaluate and choose the reliable blockchain peers, Zheng \textit{et al.} \cite{zheng2019selecting} proposed a hybrid reliability prediction model for blockchains named H-BRP, which is able to predict the reliability of blockchain peers by extracting their reliability parameters.
 
 }
 
 
 \subsection{ {\color{black} Scalability-Improving Solutions}}
 
  {\color{black}
   
  One of the critical bottlenecks of today's blockchain systems is the scalability. For example, the throughput of a blockchain is not scalable when the network size grows. To address this dilemma, a number of scalability approaches have been proposed. In this part, we conduct an overview of the most recent solutions with respect to Sharding techniques, interoperability among multiple blockchains, and other solutions.
  \modified{We summarize this subsection in Table \ref{Table:scalability}.}
  
  }
  
 
\subsubsection{{\color{black} Solutions to Sharding Blockchains}}

  {\color{black}
  
    Bitcoin's transaction throughput does not scale well. The solutions that use classical Byzantine consensus protocols do not work in an open environment like cryptocurrencies. To solve the above problems, Luu \textit{et al.} \cite{luu2016secure} proposed a new distributed agreement protocol for the permission-less blockchains, called \textit{Elastico}, which is viewed as the first secure candidate for a sharding protocol towards the open public blockchains that tolerate a constant fraction of byzantine-fault network nodes. The key idea in Elastico is to partition the network into smaller committees, each of which processes a disjoint set of transactions or a \textit{shard}. The number of committees grows linearly in the total computational power of the network. Using Elastico, the blockchain's transaction throughput increases almost linearly with the computational power of the network.

   Some early-stage sharding blockchain protocols (e.g., Elastico) improve the scalability by enforcing multiple groups of committees work in parallel. However, this manner still requires a large amount of communication for verifying every transaction linearly increasing with the number of nodes within a committee. Thus, the benefit of sharding policy was not fully employed. As an improved solution, Zamani \textit{et al.} \cite{zamani2018rapidchain} proposed a Byzantine-resilient sharding-based protocol, namely Rapidchain, for permissionless blockchains. Taking the advantage of block pipelining, RapidChain improves the throughput by using a sound intra-committee consensus. The authors also developed an efficient cross-shard verification method to avoid the broadcast messages flooding in the holistic network.

 To enforce the throughput scaling with the network size, Gao \textit{et al.}  \cite{gao2019jaciii} proposed a scalable blockchain protocol, which leverages both sharding and Proof-of-Stake consensus techniques. Their experiments were performed in an Amazon EC2-based simulation network. Although the results showed that the throughput of the proposed protocol increases following the network size, the performance was still not so high, for example, the maximum throughput was 36 transactions per second and the transaction latency was around 27 seconds.

 Aiming to improve the efficiency of cross-shard transactions, Amiri \textit{et al.} \cite{amiri2019sharper} proposed a permissioned blockchain system named \textit{SharPer}, which is strive for the scalability of blockchains by dividing and reallocating different data shards to various network clusters. 
 The major contributions of the proposed SharPer include the related algorithm and protocol associated to such SharPer model. In the author's previous work, they have already proposed a permissioned blockchain, while in this paper the authors extended it by introducing a consensus protocol in the processing of both intra-shard and cross-shard transactions. Finally, SharPer was devised by adopting sharding techniques. One of the important contributions is that SharPer can be used in the networks where there are a high percentage of non-faulty nodes. Furthermore, this paper also contributes a flattened consensus protocol w.r.t the order of cross-shard transactions among all involved clusters.

 Considering that the Ethereum places each group of transactions on a shard by their account addresses, the workloads and complexity of transactions in shards are apparently unbalanced. This manner further damages the network throughput. To address this uneven problem, Kim \textit{et al.} \cite{kim2019gas} proposed D-GAS, which is a dynamic load balancing mechanism for Ethereum shards. Using such D-GAS, the transaction workloads of accounts on each shard can be reallocated according to their gas consumption. The target is to maximize the throughput of those transactions. The evaluation results showed that the proposed D-GAS achieved at most a 12\% superiority of transaction throughput and a 74\% lower transaction latency comparing with other existing techniques.

}


{\color{black}

 The random sharding strategy causes imbalanced performance gaps among different committees in a blockchain network. Those gaps yield a bottleneck of transaction throughput. Thus, Wang \textit{et al.} \cite{Wang2019sharding} proposed a new sharding policy for blockchains named NRSS, which exploits node rating to assess network nodes according to their performance of transaction verifications. After such evaluation, all network nodes will be reallocated to different committees aiming at filling the previous imbalanced performance gaps. Through the experiments conducted on a local blockchain system, the results showed that NRSS improves throughput by around 32\% under sharding techniques.
 
 }

{\color{black}

Sharding has been proposed to mainly improve the scalability and the throughput performance of blockchains. A good sharding policy should minimize the cross-shard communications as much as possible. A classic design of sharding is the \textit{Transactions Sharding}. However, such Transactions Sharding exploits the \textit{random sharding} policy, which leads to a dilemma that most transactions are cross-shard. To this end, Nguyen \textit{et al.} \cite{nguyen2019optchain} proposed a new sharding paradigm differing from the random sharding, called OptChain, which can minimize the number of cross-shard transactions. The authors achieved their goal through the following two aspects. First they designed two metrics, named  T2S-score (Transaction-to-Shard) and L2S-score (Latency-to-Shard), respectively. T2S-score aims to measure how likely a transaction should be placed into a shard, while L2S-score is used to measure the confirmation latency when placing a transaction into a shard. Next, they utilized a well-known PageRank analysis to calculate T2S-score and proposed a mathematical model to estimate L2S-score. Finally, how does the proposed OptChain place transactions into shards based on the combination of T2S and L2S scores? In brief, they introduced another metric composed of both T2S and L2S, called  \textit{temporal fitness} score. For a given transaction $u$ and a shard $S_i$, OptChain figures the temporal fitness score for the pair $\langle u, S_i \rangle$. Then, OptChain just puts transaction $u$ into the shard that is with the highest temporal fitness score.

 Similar to \cite{nguyen2019optchain}, Dang \textit{et al.} \cite{dang2019towards} proposed a new shard-formation protocol, in which the nodes of different shards are re-assigned into different committees to reach a certain safety degree. In addition, they also proposed a coordination protocol to handle the cross-shard transactions towards guarding against the Byzantine-fault malicious coordinators. The experiment results showed that the throughput achieves a few thousands of TPS in both a local cluster with 100 nodes and a large-scale Google cloud platform testbed.
 
 Considering that the reshuffling operations lead to huge data migration in the sharding-based protocols, Chen \textit{et al.} \cite{chen2019sschain} devised a non-reshuffling structure called SSChain. Such new sharding-based protocol can avoid the overhead of data migration while enabling both transaction sharding and state sharding. Their evaluation results showed that SSChain achieves at least 6500 TPS in a network with 1800 nodes and no periodical data-migration needed.

  Multiple chains can help increase the throughput of the blockchain. However, one issue under multiple-chain system must be solved. That is, the logical ordering of blocks generated should be guaranteed, because the correct logical order is critical to the confirmation of transactions. To this end, Niu \textit{et al.} \cite{niu2019eunomia} proposed Eumonia, which is a permissionless parallel-chain protocol towards a global ordering of blocks. The authors implemented Eunomia by exploiting a fine-grained UTXO sharding model, in which the conflicted transactions can be well handled, and such protocol is proved as Simple Payment Verification (SPV) friendly.

 Although the sharding techniques have received much interests recently, it should be noticed that the committee organization is easily to attract Sybil attacks, in which a malicious node can compromise the consensus by creating multiple dummy committee members in the vote phase of the consensus protocol. To address such Sybil attacks, Rajab \textit{et al.} \cite{rajab2020feasibility} systematically formulated a model and performed an analysis w.r.t the vulnerability of Sybil attacks in the pioneer sharding protocol Elastico \cite{luu2016secure}. The authors found that the blockchain nodes that have high hash-computing power are capable to manipulate Elastico protocol using a large number of Sybil IDs. The other two conditions of Sybil attacks were derived and evaluated by numerical simulations.

 The traditional Sharding blockchain protocols can only endure up to 1/3 Byzantine-fault nodes within a shard. This weak BFT feature makes the number of nodes inside a shard cannot be small to ensure the shard functions securely. To improve the sustainability of blockchain sharding, Xu \textit{et al.} \cite{xu2020n}  proposed a new BFT sharding approach that can tolerate at most 1/2 Byzantine-fault nodes existing inside a shard. This approach benefits the throughput of decentralized databases.

 Although the existing sharding-based protocols, e.g., Elastico, OminiLedger and RapaidChain, have gained a lot of attention, they still have some drawbacks. For example, the mutual connections among all honest nodes require a big amount of communication resources. Furthermore, there is no an incentive mechanism driven nodes to participate in sharding protocol actively. To solve those problems, Zhang \textit{et al.}  \cite{zhang2020cycledger} proposed \textit{CycLedger}, which is a protocol designed for the sharding-based distributed ledger towards scalability, reliable security, and incentives. Such the proposed CycLedger is able to select a leader and a subset of nodes for each committee that handle the intra-shard consensus and the synchronization with other committees. A semi-commitment strategy and a recovery processing scheme were also proposed to deal with system crashing. In addition, the authors also proposed a reputation-based incentive policy to encourage nodes behaving honestly. 

}

 \subsubsection{ {\color{black} Multiple-Chain \& Cross-Chain: Interoperability amongst Multiple Blockchains}}

\begin{figure}[t]
	\centering
	\vspace{-3mm}
	\includegraphics[width=0.5\textwidth]{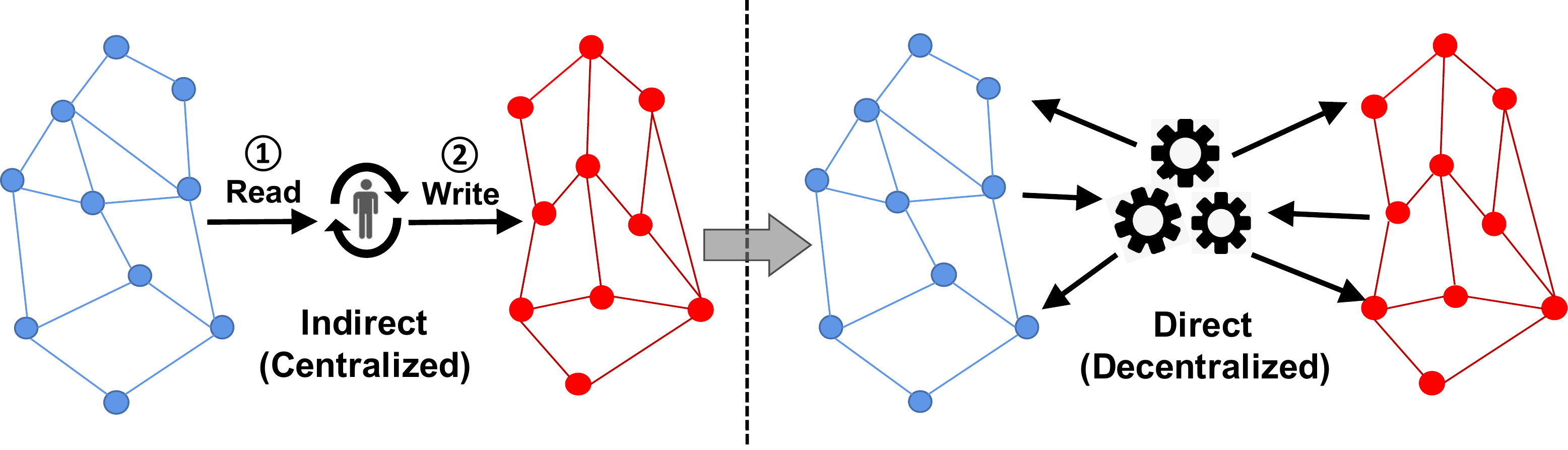}
	\vspace{-2mm}
	\caption{\modified{The illustration of interoperability across blockchains \cite{jin2018towards}. The left figure demonstrates the indirect way of interoperability that requires a centralized third party. The right figure demonstrates the direct way of interoperability without the presence of any third party.}}
	\label{fig:interoperabilityMap}
	\vspace{-3mm}
\end{figure}
\begin{figure}[t]
	\centering
	\includegraphics[width=0.5\textwidth]{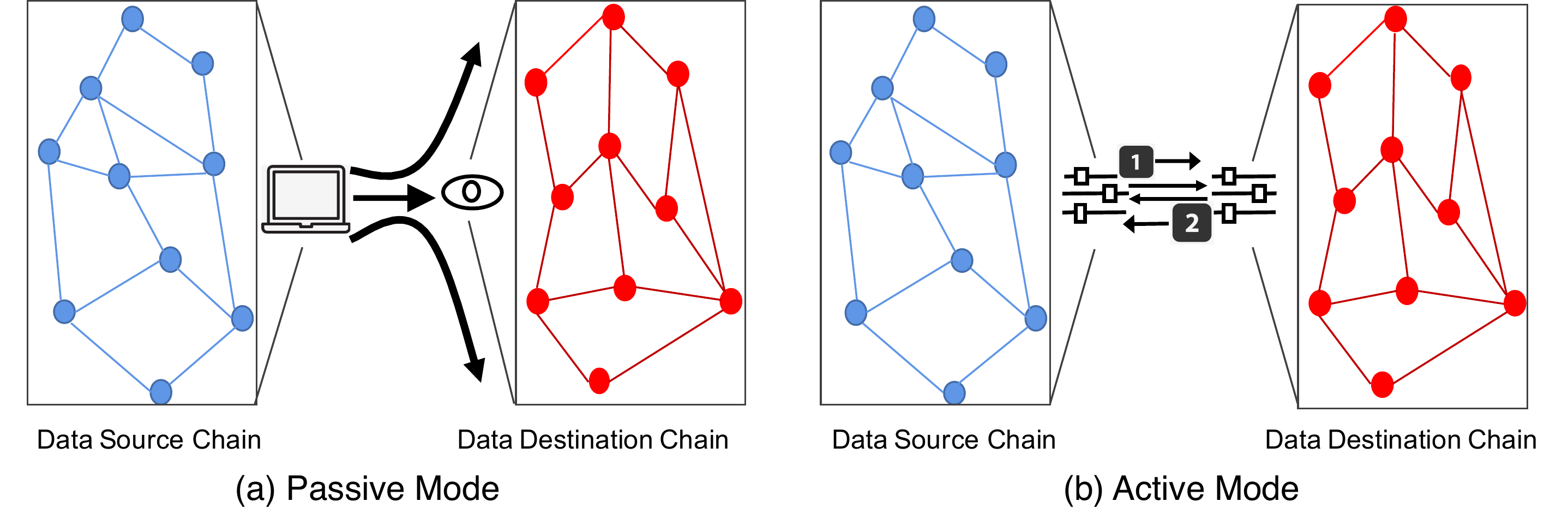}
	\vspace{-2mm}
	\caption{\modified{The interoperability of blockchains \cite{jin2018towards}. Passive mode is shown in the left figure, in which the source chain is monitored by the destination chain instead of actively sending information to the destination chain as shown in the right figure.}}
	\label{fig:interoperability}
	\vspace{-3mm}
\end{figure}

  {\color{black}
  
  The interoperability of blockchains plays a significant role for the cross-chain transactions. Such interoperability mainly includes  the effective communications and data exchange amongst multiple blockchains, as shown in Fig. \ref{fig:interoperabilityMap}. A lot of theoretical and practical issues of this direction need urgent solutions. Some representative studies are reviewed as follows.

 To enable rich functionalities and capabilities for the future blockchain ecosystems, Jin \textit{et al.} \cite{jin2018towards} proposed a novel interoperability architecture that supports the cross-chain cooperation among multiple blockchains, such as bitcoin and Ethereum. The authors classified the interoperability of multiple-chain ecosystems into passive and active modes, which are shown in Fig. \ref{fig:interoperability}. Then, the authors introduced a particular method, called Monitor Multiplexing Reading (MMR), dedicated to the passive cross-chain communications.

  Following the widespread adoption of smart contracts, the roles of blockchains have been upgraded from token exchanges into programmable state machines. Thus, the blockchain interoperability must evolve accordingly. To help realize such new type of interoperability among multiple heterogeneous blockchains, Liu \textit{et al.}\cite{liu2019hyperservice} proposed HyperService, which includes two major components, i.e., a programming framework allowing developers to create cross-chain applications; and a universal interoperability protocol towards secure implementation of DApps on blockchains. The authors implemented a 35,000-line prototype to prove the practicality of HyperService. Using the prototype, the end-to-end delays of cross-chain DApps, and the aggregated platform throughput can be measured conveniently.

In an ecosystem that consists of multiple blockchains, interoperability among those difference blockchains is an essential issue. To help the smart-contract developers build DApps, Fynn \textit{et al.} \cite{fynn2020smart} proposed a practical \textit{Move} protocol that works for multiple blockchains. The basic idea of such protocol is to support a move operation enabling to move objects and smart contracts from one blockchain to another.
Recently, to enable cross-cryptocurrency transactions, Tian \textit{et al.} \cite{tian2020enabling} proposed a decentralized cryptocurrency exchange strategy implemented on Ethereum through smart contracts.
  Additionally, a great number of studies of cross-chain communications are included in \cite{zamyatin2019sok},  in which readers can find a systematic classification of cross-chain communication protocols.
 
 }

 \subsection{{\color{black} New Protocols and Infrastructures}}
\modified{This subsection is summarized in Table \ref{Table:newProto}.}
\subsubsection{{\color{black} New Protocols for Blockchains}}

  {\color{black}
  
 David \textit{et al.} \cite{david2018ouroboros} proposed a provably secure PoS protocol named \textit{Ouroboros Praos}, which particularly exploits forward secure digital signatures and a verifiable random function such that the proposed Ouroboros Praos can endure any corruption towards any participants from an adversary in a given message delivery delay.


In blockchain systems, a node only connects to a small number of neighbor nodes. Mutual communications are achieved by gossip-like P2P messages. Based on such P2P gossip communications,  Buchman \textit{et al.} \cite{buchman2018latest} proposed a new protocol named Tendermint, which serves as a new termination mechanism for simplifying BFT consensus protocol.

 In Monoxide proposed by \cite{wang2019monoxide}, the authors have devised a novel proof-of-work scheme, named \textit{Chu-ko-nu mining}. This new proof protocol encourages a miner to create multiple blocks in different zones simultaneously with a single PoW solving effort. This mechanism makes the effective mining power in each zone is almost equal to the level of the total physical mining power in the entire network. Thus, Chu-ko-nu mining increases the attack threshold for each zone to 50\%. Furthermore, Chu-ko-nu mining can improve the energy consumption spent on mining new blocks because a lot of more blocks can be produced in each round of normal PoW mining.

 The online services of crowdsourcing face a challenge to find a suitable consensus protocol. By leveraging the advantages of the blockchain such as the traceability of service contracts, Zou \textit{et al.} \cite{zou2018proof} proposed a new consensus protocol, named \textit{Proof-of-Trust} (PoT) consensus, for crowdsourcing and the general online service industries. Basically, such PoT consensus protocol leverages a trust management of all service participants, and it works as a hybrid blockchain architecture in which a consortium blockchain integrates with a public service network.
  
  }

\begin{table*}[t]
\caption{New Protocols \& Infrastructures to Improving the Performance of Blockchains.}
\centering
\footnotesize
\begin{tabular}{|p{0.1\textwidth}|p{0.03\textwidth}|p{0.14\textwidth}|p{0.6\textwidth}|}%
\hline
\textbf{Emphasis} &\textbf{Ref.} &\textbf{Recognition} &\textbf{Methodology}\\
\hline

	%
	\multirow{6}*{New  Protocols}
	&  \cite{david2018ouroboros}    & Ouroboros Praos & Authors proposed a new secure Proof-of-stake protocol named \textit{Ouroboros Praos}, which is proved secure in the semi-synchronous adversarial setting.\\
	
	\cline{2-4}
	{ } &  \cite{buchman2018latest}  & Tendermint & Authors proposed a new BFT consensus protocol for the wide area network organized by the gossip-based P2P network under adversarial conditions.\\

	\cline{2-4}
	{ } &  \cite{wang2019monoxide}  & Chu-ko-nu mining & Authors proposed a novel proof-of-work scheme, named \textit{Chu-ko-nu mining}, which incentivizes miners to create multiple blocks in different zones with only a single PoW mining.\\

	\cline{2-4}
	{ } &  \cite{zou2018proof}  & Proof-of-Trust (PoT)  & Authors proposed a novel Proof-of-Trust consensus for the online services of crowdsourcing.\\
	
	\hline
	\multirow{10}*{New} 
	& \cite{istvan2018streamchain} & StreamChain  & Authors proposed to shift the block-based distributed ledgers to a new paradigm of \textit{stream transaction processing} to achieve a low end-to-end latencies without much affecting throughput.\\
	
	\cline{2-4}
	\multirow{6}*{Infrastructures }
	&  \cite{amiri2019caper} & CAPER: Cross-App Trans. handling  & Authors proposed a permissioned blockchain named CAPER that can well manage both the internal and the cross-application transactions for distributed applications.\\
	
	\cline{2-4}
	\multirow{4}*{\& Architectures} 
	&   \cite{chang2020incentive} & Optimal mining for miners  & Authors proposed an edge computing-based blockchain network architecture, aiming to allocate optimal computational resources for miners.\\

	\cline{2-4}
	{ } &   \cite{zheng2020axechain} & AxeChain: Useful Mining  & Authors proposed a new framework for practical PoW blockchains called AxeChain, which can spend computing power of blockchains to solve arbitrary practical problems submitted by system clients.\\
	 
	 \cline{2-4}
	{ } &   \cite{chen2020nonlinear}  & Non-linear blockchain system  & Authors explored three major metrics of blockchains, and devised a non-linear blockchain system.\\
	  
	\hline
	
\end{tabular}
\label{Table:newProto}
\end{table*}

\subsubsection{{\color{black} New Infrastructures \& Architectures for Blockchains}}

  {\color{black}
  
 Conventionally, block-based data structure is adopted by permissionless blockchain systems as blocks can efficiently amortize the cost of cryptography. However, the benefits of blocks are saturated in today's permissioned blockchains since the block-processing introduces large batching latencies. To the distributed ledgers that are neither geo-distributed nor Pow-required, Istv{\'a}n \textit{et al.}  \cite{istvan2018streamchain}  proposed to shift the traditional block-based data structure into the paradigm of \textit{stream-like transaction processing}. The premier advantage of such paradigm shift is to largely shrink the end-to-end latencies for permissioned blockchains. The authors developed a prototype of their concept based on Hyperledger Fabric. The results showed that the end-to-end latencies achieved sub-10 ms and the throughput was close to 1500 TPS.

 Permissioned blockchains have a number of limitations, such as poor performance, privacy leaking, and inefficient cross-application transaction handling mechanism. To address those issues, Amiri \textit{et al.} \cite{amiri2019caper} proposed CAPER, which a permissioned blockchain that can well deal with the cross-application transactions for distributed applications. In particular, CAPER constructs its blockchain ledger using DAG and handles the cross-application transactions by adopting three specific consensus protocols, i.e., a global consensus using a separate set of orders, a hierarchical consensus protocol, and a \textit{one-level} consensus protocol.
 %
 %
  %
  Then, Chang \textit{et al.} \cite{chang2020incentive} proposed an edge computing-based blockchain \cite{li2019credit} architecture, in which edge-computing providers supply computational resources for blockchain miners. The authors then formulated a two-phase stackelberg game for the proposed architecture, aiming to find the Stackelberg equilibrium of the theoretical optimal mining scheme.
 %
  Next, Zheng \textit{et al.} \cite{zheng2020axechain} proposed a new infrastructure for practical PoW blockchains called AxeChain, which aims to exploit the precious computing power of miners to solve arbitrary practical problems submitted by system users. The authors also analyzed the trade-off between energy consumption and security guarantees of such AxeChain. This study opens up a new direction for pursing high energy efficiency of meaningful PoW protocols.
 
 With the non-linear (e.g., graphical) structure adopted by blockchain networks, researchers are becoming interested in the performance improvement brought by new data structures. To find insights under such non-linear blockchain systems, Chen \textit{et al.} \cite{chen2020nonlinear} performed a systematic analysis by taking three critical metrics into account, i.e., \textit{full verification}, \textit{scalability}, and \textit{finality-duration}. The authors revealed that it is impossible to achieve a blockchain that enables those three metrics at the same time. Any blockchain designers must consider the trade-off among such three properties.  
 
 }

 
 \section{Various Modelings and Techniques for Better Understanding Blockchains}\label{sec:understand}
\modified{We summarize various analytical models for blockchain networks in Table \ref{Table:modelings} and Table \ref{Table:analytics}.}
 
 \subsection{{\color{black} Graph-based Theories}}

  {\color{black}
  
 The graphs are widely used in blockchain networks. For example, Merkel Tree has been adopted by Bitcoin, and several blockchain protocols, such as Ghost \cite{sompolinsky2015secure}, Phantom \cite{sompolinsky2018phantom}, and Conflux \cite{li2018scaling}, constructed their blocks using  the directed acyclic graph (DAG) technique.
 Different from those generalized graph structures, we review the most recent studies that exploit the graph theories for better understanding blockchains in this part.

  Since the transactions in blockchains are easily structured into graphs, the graph theories and graph-based data mining techniques are viewed as good tools to discover the interesting findings beyond the graphs of blockchain networks. 
  Some representative recent studies are reviewed as follows. 
  
Leveraging the techniques of graph analysis, Chen \textit{et al.} \cite{chen2018understanding} characterized three major activities on Ethereum, i.e., money transfer, the creation of smart contracts, and the invocation of smart contracts. The major contribution of this paper is that it performed the first systematic investigation and proposed new approaches based on cross-graph analysis, which can address two security issues existing in Ethereum: attack forensics and anomaly detection.
Particularly, w.r.t the graph theory, the authors mainly concentrated on the following two aspects:
\begin{enumerate}
     \item \textit{Graph Construction}: They identified four types of transactions that are not related to money transfer, smart contract creation, or smart contract invocation.
     \item \textit{Graph Analysis}: Then, they divided the remaining transactions into three groups according to the activities they triggered, i.e., money flow graph (MFG), smart contract creation graph (CCG) and contract invocation graph (CIG).
 \end{enumerate}
 Via this manner, the authors delivered many useful insights of transactions that are helpful to address the security issues of Ethereum.
 %
 Similarly, by processing Bitcoin transaction history, Akcora \textit{et al.} \cite{akcora2018bitcoin} and Dixon \textit{et al.} \cite{dixon2019blockchain} modeled the transfer network into an extreme transaction  graph. Through the analysis of chainlet  activities \cite{akcora2018forecasting}  in the constructed graph, they proposed to use GARCH-based forecasting models to identify the financial risk of Bitcoin market for cryptocurrency users.

 An emerging research direction associated with blockchain-based cryptocurrencies is to understand the network dynamics behind graphs of those blockchains, such as the transaction graph. This is because people are wondering what the connection between the price of a cryptocurrency and the dynamics of the overlying transaction graph is. To answer such a question, Abay \textit{et al.} \cite{abay2019chainnet}  proposed Chainnet, which is a computationally lightweight method to learning the graph features of blockchains. The authors also disclosed several insightful findings. For example, it is the topological feature of transaction graph that impacts the prediction of Bitcoin price dynamics, rather than the degree distribution of the transaction graph.

  Furthermore, utilizing the Mt. Gox transaction history, Chen \textit{et al.} \cite{chen2019market} also exploited the graph-based data-mining approach to dig the market manipulation of Bitcoin. The authors constructed three graphs, i.e., extreme high graph (EHG), extreme low graph (ELG), and normal graph (NMG), based on the initial processing of transaction dataset. Then, they discovered many correlations between market manipulation patterns and the price of Bitcoin.

 On the other direction, based on \textit{address graphs}, Victor \textit{et al.} \cite{victor2019measuring} studied the ERC20 token networks through analyzing smart contracts of Ethereum blockchain.  Different from other graph-based approaches, the authors focused on their attention on the address graphs, i.e., token networks. With all network addresses, each token network is viewed as an overlay graph of the entire Ethereum network addresses. Similar to \cite{chen2018understanding}, the authors presented the relationship between transactions by exploiting graph-based analysis, in which the arrows can denote the invoking functions between transactions and smart contracts, and the token transfers between transactions as well. The findings presented by this study help us have a well understanding of token networks in terms of time-varying characteristics, such as the usage patterns of the blockchain system. An interesting finding is that around 90\% of all transfers stem from the top 1000 token contracts. That is to say, only less than 10\% of  token recipients have transferred their tokens. This finding is contrary to the viewpoint proposed by \cite{somin2018network}, where Somin \textit{et al.} showed that the full transfers seem to obey a power-law distribution. However, the study  \cite{victor2019measuring} indicated that those transfers in token networks likely do not follow a power law. The authors attributed such the observations to the following three possible reasons: 1) most of the token users don't have incentives to transfer their tokens. Instead, they just simply hold tokens; 2) the majority of inactive tokens are treated as something like unwanted spam; 3) a small portion, i.e., approximately 8\%, of users intended to sell their tokens to a market exchange.

 Recently, Zhao \textit{et al.} \cite{zhao2020exploring} explored the account creation, account vote, money transfer and contract authorization activities of early-stage EOSIO transactions through graph-based metric analysis. Their study revealed abnormal transactions like voting gangs and frauds.
 }

\subsection{{\color{black} Stochastic Modelings}}

 {\color{black}

 The latencies of block transfer and processing are generally existing in blockchain networks since the large number of miner nodes are geographically distributed. Such delays increase the probability of forking and the vulnerability to malicious attacks. Thus, it is critical to know how would the network dynamics caused by the block propagation latencies and the fluctuation of hashing power of miners impact the blockchain performance such as block generation rate. To find the connection between those factors,  Papadis \textit{et al.} \cite{papadis2018stochastic} developed stochastic models to derive the blockchain evolution in a wide-area network. Their results showed us practical insights for the design issues of blockchains, for example, how to change the difficulty of mining in the PoW consensus while guaranteeing an expected block generation rate or an immunity level of adversarial attacks. The authors then performed analytical studies and simulations to evaluate the accuracy of their models. This stochastic analysis opens up a door for us to have a deeper understanding of dynamics in a blockchain network.
 

Towards the stability and scalability of blockchain systems, Gopalan \textit{et al.} \cite{gopalan2020stability} also proposed a stochastic model for a blockchain system. During their modeling, a structural asymptotic property called \textit{one-endedness} was identified. The authors also proved that a blockchain system is one-ended if it is stochastically stable. The upper and lower bounds of the stability region were also studied. The authors found that the stability bounds are closely related to the conductance of the P2P blockchain network. Those findings are very insightful such that researchers can assess the scalability of blockchain systems deployed on large-scale P2P networks.

 Although Sharding protocol is viewed as a very promising solution to solving the scalability of blockchains and adopted by multiple well-known blockchains such as RapidChain \cite{zamani2018rapidchain}, OmniLedger \cite{kokoris2018omniledger}, and Monoxide \cite{wang2019monoxide}, the failure probability for a committee under Sharding protocol is still unknown. To fill this gap, Hafid \textit{et al.} \cite{Hafid2019problistic, hafid2019methodology, hafid2019new} proposed a stochastic model to capture the security analysis under Sharding-based blockchains using a probabilistic approach. With the proposed mathematical model, the upper bound of the failure probability was derived for a committee. In particular, three probability inequalities were used in their model, i.e., Chebyshev, Hoeffding, and Chv{\'a}tal. The authors claim that the proposed stochastic model can be used to analyze the security of any Sharding-based protocol.
 
 }

\begin{table*}[t]
\caption{Various Modelings, Techniques and Theories for Better Understanding Blockchains.}
\centering
\footnotesize
\begin{tabular}{|p{0.07\textwidth}|p{0.11\textwidth}|p{0.07\textwidth}|p{0.21\textwidth}|p{0.38\textwidth}|}%
\hline
\textbf{Category}&\textbf{Emphasis} &\textbf{Ref.} &\textbf{Metrics} &\textbf{Methodology \& Implications}\\
\hline


	\multirow{12}*{Graph-} 
	& \multirow{8}*{Transactions} &  \cite{chen2018understanding} &  Cross-graph analysis of Ethereum & Via graph analysis, authors extracted three major activities, i.e., money transfer, smart contracts creation, and  smart contracts invocation.\\

	\cline{3-5}
	
	\multirow{8}*{based} 
	&\multirow{4}*{mining}& \cite{abay2019chainnet} & Features of transaction graphs  & Proposed an extendable and computationally efficient method for graph representation learning on Blockchains.\\
	 
	\cline{3-5}
	
	\multirow{6}*{Theories} 
	&{}& \cite{chen2019market} & Market manipulation
patterns  & Authors exploited the graph-based data-mining approach to reveal the market manipulation evidence of Bitcoin.\\
	 
	\cline{3-5}
	{ }&{}&  \cite{zhao2020exploring} & Clustering coefficient, assortativity of TX graph & Authors exploited the graph-based analysis to reveal the abnormal transactions of EOSIO.\\

	\cline{2-5}
	{ }&\multirow{5}*{Token networks}& \cite{victor2019measuring} & Token-transfer distributions  & Authors studied the token networks through analyzing smart contracts of Ethereum blockchain based on graph analysis.\\
	 
	\cline{3-5}
	{ }&{}& \cite{akcora2018bitcoin, dixon2019blockchain} & Extreme chainlet activity & Authors proposed graph-based analysis models for assessing the financial investment risk of \modified{Bitcoin}.\\
	 
	\hline


	\multirow{12}*{Stochastic} 
	 & {Blockchain network analysis} & \cite{papadis2018stochastic} & Block completion rates, and the probability of a successful adversarial attack  & Authors derived stochastic models to capture critical blockchain properties, and to evaluate the impact of blockchain propagation latency on key performance metrics. This study provides us useful insights of design issues of blockchain networks.\\

	\cline{2-5}
	\multirow{4}*{Modelings} 
	& \multirow{3}*{Stability analysis} & \multirow{1}*{\cite{gopalan2020stability} }  & Time to consistency, cycle length, consistency fraction, age of information   & Authors proposed a network model which can identify the stochastic stability of blockchain systems. 
	\\

	\cline{2-5}
	{} & Failure probability analysis & \cite{Hafid2019problistic, hafid2019methodology, hafid2019new}  & Failure probability of a committee, sums of upper-bounded hypergeometric and binomial distributions for each epoch  & Authors proposed a probabilistic model to derive the security analysis under Sharding blockchain protocols. This study can tell how to keep the failure probability smaller than a defined threshold for a specific sharding protocol.\\
	 
	\hline


	\multirow{14}*{Queueing} 
	& Mining procedure and block-generation &  \cite{li2018blockchain, li2019markov} & The average number of TX in the arrival queue and in a block, and average confirmation time of TX  & Authors developed a \modified{Markovian} batch-service queueing system to express the mining process and the generation of new blocks in miners pool.\\
  
	\cline{2-5}
	\multirow{10}*{Theories} 
	& Block-confirmation time & \cite{ricci2019learning} & The residual lifetime of a block till the next block is confirmed & Authors proposed a theoretical framework to deeply understand the transaction confirmation time, by  integrating the queueing theory and machine learning techniques. \\

	\cline{2-5}
	& Synchronization process of Bitcoin network & \cite{frolkova2019bitcoin} &  Stationary queue-length distribution & Authors proposed an infinite-server model with random fluid limit for Bitcoin network.\\
 
	\cline{2-5}
	& Mining resources allocation &  \cite{fang2020toward} & Mining resource for miners, queueing stability & Authors proposed a Lyapunov optimization-based queueing analytical model to study the allocation of mining resources for the PoW-based blockchain networks. \\
	
 	\cline{2-5}
	& Blockchain's theoretical working principles & \cite{memon2019simulation} & number of TX per block, mining interval of each block, memory pool size, waiting time, number of unconfirmed TX & Authors proposed a queueing theory-based model to have a better understanding the theoretical working principle of blockchain networks. \\

	\hline
	
\end{tabular}
\label{Table:modelings}
\end{table*}

 \subsection{{\color{black} Queueing Theories for Blockchain Systems}}
 
 {\color{black}
 
 In blockchain networks, several stages of mining processing and the generation of new blocks can be formulated as queueing systems, such as the transaction-arrival queue, the transaction-confirmation queue, and the block-verification queue. Thus, a growing number of studies are exploiting the queueing theory to disclose the mining and consensus mechanisms of  blockchains. Some recent representative works are reviewed as follows.

 To develop a queueing theory of blockchain systems,  Li \textit{et al.} \cite{li2018blockchain, li2019markov} devised a batch-service queueing system to describe the mining and the creating of new blocks in miners' pool. For the blockchain queueing system, the authors exploited the type GI/M/1 continuous-time Markov process. Then, they derived the stable condition and the stationary probability matrix of the queueing system utilizing the matrix-geometric techniques.
  
  Then, viewing that the confirmation delay of Bitcoin transactions are larger than conventional credit card systems, Ricci \textit{et al.} \cite{ricci2019learning} proposed a theoretical framework integrating the queueing theory and machine learning techniques to have a deep understanding towards the transaction confirmation time. The reason the authors chose the queueing theory for their study is that a queueing model is suitable to see insights into how the different blockchain parameters affect the transaction latencies. Their measurement results showed that the Bitcoin users experience a delay that is slightly larger than the residual time of a block confirmation.
  
  Frolkova \textit{et al.} \cite{frolkova2019bitcoin} formulated the synchronization process of Bitcoin network as an infinite-server model. The authors derived a closed-form for the model that can be used to capture the queue stationary distribution. Furthermore, they also proposed a random-style fluid limit under service latencies.

  On the other hand, to evaluate and optimize the performance of blockchain-based systems,  Memon  \textit{et al.} \cite{memon2019simulation} proposed a simulation model by exploiting queueing theory. In the proposed model, the authors constructed an  M/M/1 queue for the memory pool, and an M/M/c queue for the mining pool, respectively. This model can capture multiple critical statistics metrics of blockchain networks, such as the number of transactions every new block, the mining interval of a block, transactions throughput, and the waiting time in memory pool, etc.
    
  Next, Fang \textit{et al.} \cite{fang2020toward} proposed a queueing analytical model to allocate mining resources for the general PoW-based blockchain networks. The authors formulated the queueing model using Lyapunov optimization techniques. Based on such stochastic theory, a dynamic allocation algorithm was designed to find a trade-off between mining energy and queueing delay. Different from the aforementioned work \cite{li2018blockchain, li2019markov, ricci2019learning}, the proposed Lyapunov-based algorithm does not need to make any statistical assumptions on the arrivals and services.
  
 }

\begin{table*}[t]
\caption{Various Analytics Models for Better Understanding Blockchain Networks.}
\centering
\footnotesize
\begin{tabular}{|p{0.15\textwidth}|p{0.03\textwidth}|p{0.2\textwidth}|p{0.48\textwidth}|}%
\hline
\textbf{Emphasis} &\textbf{Ref.} &\textbf{Metrics} &\textbf{Methodology \& Implications}\\
\hline


	%
  	\multirow{4}*{Applicability}
	 &  \cite{wust2018you} &  Public verifiability, transparency, privacy, integrity, redundancy, and trust anchor   & Authors proposed the first structured analytical methodology that can help decide whether a particular application system indeed needs a blockchain, either a permissioned or permissionless, as its technical solution.\\
  
	\cline{2-4}
  	\multirow{1}*{of blockchains} & \modified{\cite{zhang2013privacy}}  & \modified{Scalability, efficiency and privacy issues in cloud for blockchains}  &  \modified{Authors proposes a novel upper bound privacy leakage based approach to identify intermediate data sets partitioned and distributed in cloud for encryption. This approach can significantly improve the scalability and efficiency of data processing for privacy preserving in cloud.}\\
  
	\hline
  	
  	\multirow{5}*{Exploration of}
	 &   \cite{lin2020modeling} &  Temporal information and the multiplicity features of Ethereum transactions  & Authors proposed an analytical model based on the multiplex network theory for understanding Ethereum transactions.\\
	
	\cline{2-4}
  	\multirow{1}*{Ethereum transactions}
  	&  \cite{sousa2019analysis} &  Pending time of Ethereum transactions  & Authors conducted a characterization study of the Ethereum by focusing on the pending time, and attempted to find the correlation between pending time and fee-related parameters of Ethereum.\\

	\hline
  	Modeling the competition over multiple miners &   \cite{altman2019blockchain} &  Competing mining resources of miners of a cryptocurrency blockchain  & Authors exploited the Game Theory to find a Nash equilibria while peers are competing mining resources.\\
  
	\hline
	A neat bound of consistency latency & \cite{zhao2020analysis}  & Consistency of a PoW blockchain  & Authors derived a neat bound of mining latencies that helps understand the consistency of Nakamoto's blockchain consensus in asynchronous networks.\\

	\hline
	Network connectivity & \cite{xiao2020modeling} &  Consensus security   & Authors proposed an analytical model to evaluate the impact of network connectivity on the consensus security of PoW blockchain under different adversary models.\\
	
	\hline
	How Ethereum responds to sharding & \cite{fynn2018challenges} &  Balance among shards,  number of TX that would involve multiple shards, the amount of data relocated across shards  & Authors studied how sharding impact Ethereum by firstly modeling Ethereum through graph modeling, and then assessing the three metrics mentioned when partitioning the graph.\\
	
	\hline
	Required properties of sharding protocols &  \cite{avarikioti2019divide} &  Consistency and Scalability   & Authors proposed an analytical model to evaluate whether a protocol for sharded distributed ledgers fulfills necessary properties.\\
	
	\hline
	Vulnerability by forking attacks & \cite{wang2019corking} &  Hashrate power, net cost of an attack  & Authors proposed fine-grained vulnerability analytical model of blockchain networks incurred by intentional forking attacks taking the advantages of large deviation theory. \\

	\hline
	Counterattack to double-spend attacks &   \cite{moroz2020doublespend} &  Robustness parameter,  vulnerability probability  & Authors studied how to defense and even counterattack the double-spend attacks in PoW blockchains.\\
	
	\hline
	Limitations of PBFT-based blockchains &  \cite{bessani2020byzantine} &  Performance of blockchain applications, Persistence, Possibility of forks  & Authors studied  and identified several misalignments between the requirements of permissioned blockchains and the classic BFT protocols.\\
 
	\hline
	\modified{Unified analysis of different PoX consensus schemes} &  \cite{yu2020unified} &  \modified{Resource sensitivity, system convergence, and resource Fairness}  & \modified{Authors proposed a new Markov model to unify the analysis of the steady-state for weighted resource distribution of different PoX-based Blockchains.} \\
 
	\hline
	
\end{tabular}
\label{Table:analytics}
\end{table*}

 \subsection{{\color{black} Analytical Models for Blockchain Networks}}
 
 \modified{This subsection is summarized in Table \ref{Table:analytics}.}
  {\color{black}
 
 
 For the people considering whether a blockchain system is needed for his/her business, a notable fact is that blockchain is not always applicable to all real-life use cases. To help analyze whether blockchain is appropriate to a specific application scenario, Wust \textit{et al.}  \cite{wust2018you} provided the first structured analytical methodology and applied it to analyzing three representative scenarios, i.e., supply chain management, interbank payments, and decentralized autonomous organizations.
 \modified{The other article \cite{zhang2013privacy} proposes a novel upper bound privacy leakage based approach to identify intermediate data sets partitioned and distributed in cloud for encryption. This approach can significantly improve the scalability and efficiency of data processing for privacy preserving in cloud. This study provides insights of scalability, efficiency and privacy issues in cloud for blockchain.}

 
 Although Ethereum has gained much popularity since its debut in 2014, the systematically analysis of Ethereum transactions still suffers from insufficient explorations. Therefore, Lin \textit{et al.} \cite{lin2020modeling} proposed to model the transactions using the techniques of multiplex network. The authors then devised several random-walk strategies for graph representation of the transactions network. This study could help us better understand the temporal data and the multiplicity features of Ethereum transactions.

To better understand the network features of an Ethereum transaction, Sousa \textit{et al.} \cite{sousa2019analysis} focused on the pending time, which is defined as the latency counting from the time a transaction is observed to the time this transaction is packed into the blockchain. The authors tried to find the correlations between such pending time with the fee-related parameters such as gas and gas price. Surprisingly, their data-driven empirical analysis results showed that the correlation between those two factors has no clear clue. This finding is counterintuitive.

 
 To achieve a consensus about the state of blockchains, miners have to compete with each other by invoking a certain proof mechanism, say PoW. Such competition among miners is the key module to public blockchains such as Bitcoin. 
 To model the competition over multiple miners of a cryptocurrency blockchain,
Altman \textit{et al.}  \cite{altman2019blockchain} exploited the Game Theory to find a Nash equilibria while peers are competing mining resources. The proposed approach help researchers well understand such competition. However, the authors also mentioned that they didn't study the punishment and cooperation between miners over the repeated games. Those open topics will be very interesting for future studies.
 
\modified{
 Besides competitions among individual miners, there are also competitions among mining pools. Malicious pools can pull off DDoS attacks to overload the victim pools' manager with invalid share submissions. The delay in verifying extra share submissions potentially impairs the hash power of the victim pool and thus undermines the potential reward for pool miners. Knowing that the chance of getting a reward is smaller, miners in the victim pools would migrate to another mining pools, which would further weaken the victim pools. To better understand this kind of competition, Wu \textit{et al.} \cite{wu2020survive} proposed a stochastic game-theoretic model in a two-mining-pool case. The authors used Q-learning algorithm to find the Nash equilibrium and maximize the long-term payoffs. The experiment showed that the smaller mining pool is more likely to attack the larger one. Also, mining pools tend to adopt lower attack level when the DDoS attack cost increases.
}

 
 To ensure the consistency of PoW blockchain in an asynchronous network,  Zhao \textit{et al.} \cite{zhao2020analysis} performed an analysis and derived a neat bound  around $\frac{2\mu}{\ln (\mu/\nu)}$, where $\mu + \nu = 1$, with $\mu$ and $\nu$ denoting the fraction of computation power dominated by the honest and adversarial miners, respectively. Such a neat bound of mining latencies is helpful to us to well understand the consistency of Nakamoto's blockchain consensus in asynchronous networks.


  Bitcoin's consensus security is built upon the assumption of honest-majority. Under this assumption, the blockchain system is thought secure only if the majority of miners are honest while voting towards a global consensus. Recent researches believe that network connectivity, the forks of a blockchain, and the strategy of mining are major factors that impact the security of consensus in Bitcoin blockchain. To provide pioneering concrete modelings and analysis,  Xiao \textit{et al.} \cite{xiao2020modeling} proposed an analytical model to evaluate the network connectivity on the consensus security of PoW blockchains.
To validate the effectiveness of the proposed analytical model, the authors applied it to two adversary scenarios, i.e., \textit{honest-but-potentially-colluding}, and \textit{selfish mining} models.

 
 Although Sharding is viewed as a prevalent technique for improving the scalability to blockchain systems, several essential questions are: what we can expect from and what price is required to pay for introducing Sharding technique to Ethereum? To answer those questions, Fynn \textit{et al.} \cite{fynn2018challenges} studied how sharding works for Ethereum by modeling Ethereum into a graph. Via partitioning the graph, they evaluated the trade-off between the edge-cut and balance. Several practical insights have been disclosed. For example, three major components, e..g, computation, storage and bandwidth, are playing a critical role when partitioning Ethereum; A good design of incentives is also necessary for adopting sharding mechanism.

 As mentioned multiple times, sharding technique is viewed as a promising solution to improving the scalability of blockchains. However, the properties of a sharded blockchain under a fully adaptive adversary are still unknown. To this end, Avarikioti \textit{et al.}  \cite{avarikioti2019divide} defined the \textit{consistency} and \textit{scalability} for sharded blockchain protocol. The limitations of security and efficiency of sharding protocols were also derived. Then, they analyzed these two properties on the context of multiple popular sharding-based protocols such as  \textit{OmniLedger},  \textit{RapidChain},  \textit{Elastico}, and  \textit{Monoxide}. Several interesting conclusions have been drawn. For example,  the authors thought that Elastico and Momoxide failed to guarantee the balance between consistency and scalability properties, while OmniLedger and RapidChain fulfill all requirements of a robust sharded blockchain protocol.

 
 Forking attacks has become the normal threats faced by the blockchain market. The related existing studies mainly focus on the detection of such attacks through transactions. However, this manner cannot prevent the forking attacks from happening. To resist the forking attacks, Wang \textit{et al.} \cite{wang2019corking} studied the fine-grained vulnerability of blockchain networks caused by intentional forks using the large deviation theory. This study can help set the robustness parameters for a blockchain network since the vulnerability analysis provides the correlation between robust level and the vulnerability probability. In detail, the authors found that it is much more cost-efficient to set the robust level parameters than to spend the computational capability used to lower the attack probability.

 
 The existing economic analysis \cite{budish2018economic} reported that the attacks towards PoW mining-based blockchain systems can be cheap under a specific condition when renting sufficient hashrate capability. Moroz \textit{et al.} \cite{moroz2020doublespend} studied how to defense the double-spend attacks in an interesting reverse direction. The authors found that the counterattack of victims can lead to a classic game-theoretic \textit{War of Attrition} model. This study showed us the double-spend attacks on some PoW-based blockchains are actually cheap. However, the defense or even counterattack to such double-spend attacks is possible when victims  are owning the same capacity as the attacker.

 Although BFT protocols have attracted a lot of attention, there are still a number of fundamental limitations unaddressed while running blockchain applications based on the classical BFT protocols. Those limitations include one related to low performance issues, and two correlated to the gaps between the state machine replication and blockchain models (i.e., the lack of strong persistence guarantees and the occurrence of forks).  To identify those limitations, Bessani \textit{et al.}  \cite{bessani2020byzantine} first studied them using a digital coin blockchain App called SmartCoin, and a popular BFT replication library called BFT-SMART, then they discussed how to tackle these limitations in a protocol-agnostic manner. The authors also implemented an experimental platform of permissioned blockchain, namely SmartChain. Their evaluation results showed that SmartChain can address the limitations aforementioned and significantly improve the performance of a blockchain application. 
 
 {\color{blue}
 The Nakamoto protocol is designed to solve the Byzantine Generals Problem for permissionless Blockchains. However, a general analytical model is still missing for capturing the steady-state profit of each miner against the competitors.
 To this end, Yu \textit{et al.} \cite{yu2020unified} studied the weighted resource distribution of proof-based consensus engines, referred to as Proof-of-X (PoX), in large-scale networks. The proposed Markov model attempts to unify the analysis of different PoX mechanisms considering three new unified metrics, i.e., resource sensitivity, system convergence, and resource fairness. }
 
 }

\begin{table*}[h!t]
\caption{Data Analytics for Better Understanding Cryptocurrency Blockchains.}
\centering
\footnotesize
\begin{tabular}{|p{0.14\textwidth}|p{0.05\textwidth}|p{0.19\textwidth}|p{0.5\textwidth}|}%
\hline
\textbf{Emphasis} &\textbf{Ref.} &\textbf{Metrics} &\textbf{Methodology \& Implications}\\
\hline

  	\multirow{3}*{Cryptojacking}
	 &   \multirow{2}*{\cite{tahir2019browsers}}  &  Hardware performance counters & Authors proposed a machine learning-based solution to prevent cryptojacking attacks.\\
	
	\cline{2-4}
	
  	\multirow{1}*{detection}
	&  \multirow{2}*{\cite{ning2019capjack}}  &  Various system resource utilization & Authors proposed an in-browser cryptojacking detection approach (CapJack), based on the latest CapsNet.\\
	
	\hline
	Market-manipulation mining &   \multirow{2}*{\cite{chen2019market} } &  Various graph characteristics of transaction graph & Authors proposed a mining approach using the exchanges collected from the transaction networks.\\
	
	\hline
	Predicting volatility of Bitcoin price &   \multirow{2}*{ \cite{dixon2019blockchain}}  &  Various graph characteristics of extreme chainlets & Authors proposed a graph-based analytic model to predict the intraday financial risk of Bitcoin market.\\

	\hline
	Money-laundering detection &    \multirow{2}*{\cite{hu2019characterizing}}  &  Various graph characteristics of transaction graph & Authors exploited machine learning models to detect potential money laundering activities from Bitcoin transactions.\\
	
	\hline
	\multirow{3}*{Ponzi-scheme} &     \multirow{2}*{\cite{vasek2018analyzing}}  &   Factors that affect scam persistence & Authors analyzed the demand and supply perspectives of Ponzi schemes on Bitcoin ecosystem.\\
	
	\cline{2-4}
	\multirow{1}*{detection} 
	&  \cite{chen2018detecting, chen2019exploiting}  &  Account and code features of smart contracts & Authors detected Ponzi schemes for Ethereum based on data mining and machine learning approaches.\\
	  
	\hline
	Design problem of cryptoeconomic systems &    \multirow{3}*{\cite{laskowski2020evidence}} &  Price of XNS token, Subsidy of App developers  & Authors presented a practical evidence-based example to show how data science and stochastic modeling can be applied to designing cryptoeconomic blockchains.\\
	
	\hline
	Pricing mining hardware &  \multirow{2}*{\cite{yaish2020pricing}} &   \multirow{2}*{Miner revenue,  ASIC value}  & Authors studied the correlation between the price of mining hardware (ASIC) and the value volatility of underlying cryptocurrency.\\
	
	\hline

\end{tabular}
\label{Table:analyticsCrptocurrency}
\end{table*}

 \subsection{{\color{black} Data Analytics for Cryptocurrency Blockchains}}
 \modified{This subsection is summarized in Table \ref{Table:analyticsCrptocurrency}.}

 \subsubsection{{\color{black} Market Risks Detection}}

 {\color{black}
 
 As aforementioned,  Akcora \textit{et al.} \cite{akcora2018bitcoin} proposed a graph-based predictive model to forecast the investment risk of Bitcoin market.
  On the other hand, with the tremendously increasing price of cryptocurrencies such as Bitcoin, hackers are imminently utilizing any available computational resources to participate in mining. Thus, any web users face severe risks from the cryptocurrency-hungry hackers. For example, the \textit{cryptojacking} attacks \cite{eskandari2018first} have raised growing attention. In such type of attacks, a mining script is embedded secretly by a hacker without notice from the user. When the script is loaded, the mining will begin in the background of the system and a large portion of hardware resources are requisitioned for mining.
  To tackle the cryptojacking attacks, Tahir \textit{et al.}  \cite{tahir2019browsers} proposed a machine learning-based solution, which leverages the hardware performance counters as the critical features and can achieve a high accuracy while classifying the parasitic miners. The authors also built their approach into a browser extension towards the widespread real-time protection for web users.
Similarly, Ning \textit{et al.} \cite{ning2019capjack} proposed \textit{CapJack}, which is an in-browser cryptojacking detector based on deep capsule network (CapsNet) \cite{sabour2017dynamic} technology.

 
 As mentioned previously, to detect potential manipulation of Bitcoin market, Chen \textit{et al.} \cite{chen2019market} proposed a graph-based mining to study the evidence from the transaction network built based on Mt. Gox transaction history. The findings of this study suggests that the cryptocurrency market requires regulation.

 To predict drastic price fluctuation of Bitcoin, Dixon \textit{et al.} \cite{dixon2019blockchain} studied the impact of extreme transaction graph (ETG) activity on the intraday dynamics of the Bitcoin prices. The authors utilized chainlets \cite{akcora2018forecasting} (sub graphs of transaction graph) for developing their predictive models.
}

 \subsubsection{{\color{black}Ponzi Schemes Detection}}

 {\color{black}
 
  Ponzi scheme \cite{bartoletti2020dissecting}, as a classic scam, is taking advantages of mainstream blockchains such as Ethereum. Data mining technologies  \cite{bartoletti2018data} are widely used for detecting Ponzi schemes.
  For example, several representative studies are reviewed as follows.
  Vasek \textit{et al.} \cite{vasek2018analyzing} analyzed the demand and supply Ponzi schemes on Bitcoin ecosystem. The authors were interested at the reasons that make those Ponzi frauds succeeded in attracting victims, and the lifetime of those scams.
   To detect such Ponzi schemes towards a healthier blockchain economic environment,  Chen \textit{et al.} \cite{chen2018detecting, chen2019exploiting} proposed a machine learning-based classification model by exploiting data mining on smart contracts of Ethereum. The experimental results showed that the proposed detection model can even identify Ponzi schemes at the very beginning when those schemes are created.

  }

 \subsubsection{{\color{black}Money-Laundering Detection}}
 
 {\color{black}
 
 Although Bitcoin has received enormous attention, it is also criticized for being carried out criminal financial activities such as ponzi schemes and money laundering.
 For example, Seo \textit{et al.} \cite{seo2018money} mentioned that money laundering conducted in the underground market can be detected using the Bitcoin mixing services. However, they didn't present an essential anti-money laundering strategy in their paper. 
 In contrast, utilizing a transaction dataset collected over three years, Hu \textit{et al.} \cite{hu2019characterizing} performed in-depth detection for discovering money laundering activities on Bitcoin network. To identify the money laundering transactions from the regular ones, the authors proposed four types of classifiers based on the graph features appeared on the transaction graph, i.e., immediate neighbors, deepwalk embeddings, node2vec embeddings and decision tree-based.
 
 }


 \subsubsection{{\color{black} Portrait of Cryptoeconomic Systems}}

 {\color{black}
 

 It is not common to introduce data science and stochastic simulation modelings into the design problem of cryptoeconomic engineering. Laskowski \textit{et al.} \cite{laskowski2020evidence} presented a practical evidence-based example to show how this manner can be applied to designing cryptoeconomic blockchains.


  Yaish \textit{et al.} \cite{yaish2020pricing} discussed the relationship between the cryptocurrency mining and the market price of the special hardware (ASICs) that supports PoW consensus. The authors showed that the decreasing volatility of Bitcoin's price has a counterintuitive negative impact to the value of mining hardware. This is because miners are not financially incentivized to participate in mining, when Bitcoin becomes widely adopted thus making its volatility decrease. This study also revealed that a mining hardware ASIC could be imitated by bonds and underlying cryptocurrencies such as bitcoins.
  
  }

 
 \section{Useful Measurements, Datasets and Experiment Tools for Blockchains}\label{sec:tools}
 

\modified{Measurements are summarized in Table \ref{Table:measurements}, and datasets are summarized in Table \ref{Table:datasetFramework}.}

 \subsection{{\color{black} Performance Measurements and Datasets for Blockchains}}
 
 {\color{black}
 
 Although diverse blockchains have been proposed in recent years, very few efforts have been devoted to measuring the performance of different blockchain systems. Thus, this part reviews the representative studies of performance measurements for blockchains. The measurement metrics include throughput, security, scalability, etc.

 As a pioneer work in this direction, Gervais \textit{et al.} \cite{gervais2016security} proposed a quantitative framework, using which they studied the security and performance of several PoW blockchains, such as Bitcoin, Litecoin, Dogecoin and Ethereum. The authors focused on multiple metrics of security model, e.g., stale block rate, mining power, mining costs, the number of block confirmations, propagation ability, and the impact of eclipse attacks. They also conducted extensive simulations for the four blockchains aforementioned with respect to the impact of block interval, the impact of block size, and throughput. Via the evaluation of network parameters about the security of PoW blockchains, researchers can compare the security performance objectively, and thus help them appropriately make optimal adversarial strategies and the security provisions of PoW blockchains.

\begin{table*}[t]
\caption{Various performance measurements of blockchains.}
\centering
\footnotesize
\begin{tabular}{|p{0.03\textwidth}|p{0.15\textwidth}|p{0.22\textwidth}|p{0.48\textwidth}|}%
\hline
\textbf{Ref.}&\textbf{Target Blockchains} &\textbf{Metrics} &\textbf{Implementation / Experiments / Methodology}\\
\hline

	\multirow{4}*{\cite{wang2019monoxide}} & General mining-based blockchains, e.g., Bitcoin and Ethereum & TPS, the overheads of  cross-zone transactions, the confirmation latency of transactions, etc.   & Monoxide was implemented utilizing C++. RocksDB was used to store blocks and TX. The real-world testing system was deployed on a distributed configuration consisting of 1200 virtual machines, with each owning 8 cores and 32 GB memory. In total 48,000 blockchain nodes were exploited in the testbed.\\
	
	\hline
	
	\multirow{5}*{\cite{yang2019prism}} & \multirow{5}*{General blockchains} &  Throughput and confirmation latency, scalability under different number of clients, forking rate, and resource utilization (CPU, network bandwidth)  & Prism testbed is deployed on Amazon EC2 instances each with 16 CPU cores, 16 GB RAM, 400 GB NVMe SSD, and a 10 Gbps network interface. In total 100 Prism client instances are connected into a topology in random 4-regular graph.\\

	\hline
	\multirow{3}*{\cite{woo2020garet} }& \multirow{3}*{Ethereum} & TX throughput, the makespan of transaction latency   & The proposed GARET algorithm was measured to outperform existing techniques by up to 12\% in TX throughput, and decrease the makespan of TX latency by about 74\% under various conditions in Sharding Ethereum. \\
	
	\hline
	\multirow{5}*{\cite{gervais2016security}} & \hspace{0.15\textwidth} Bitcoin, Litecoin, Dogecoin, Ethereum &  Block interval, block size, and throughput  &  Proposed a quantitative framework, using which they studied the security and performance of several PoW blockchains. Via the evaluation of network parameters about the security of PoW blockchains, researchers can make trade-offs between the security provisions and performance objectively.\\

	\hline
	\multirow{3}*{\cite{nasir2018performance}} & \multirow{3}*{Hyperledger Fabric}  & Execution time, latency, throughput, scalability vs the number  of blockchain nodes & Presented the performance measurement and analysis towards Hyperledger Fabric version 0.6 and version 1.0.\\

	\hline
	\multirow{4}*{\cite{zheng2018detailed}} & Ethereum, Parity, CITA, Hyperledger Fabric & TPS, Average response delay, Transactions per CPU, TX per memory second, TX per disk I/O and TX per network data & Proposed a scalable framework for monitoring the real-time performance blockchain systems. The authors evaluated four popular blockchain systems, i.e., Ethereum, Parity, CITA and Hyperledger Fabric.\\

	\hline
	\multirow{5}*{\cite{dinh2017blockbench}} & \multirow{5}*{Private blockchains} &  Throughput and latency, Scalability, Fault tolerance and security, and other micro measurements, e.g., CPU utilization, Network utilization, etc.  & The authors proposed Blockbench for measuring and analyzing the multiple performance of private blockchain systems. Through this Blockbench, the authors revealed several insightful bottlenecks and trade-offs while designing the software of blockchains.\\
	 
	\hline
	\multirow{3}*{\cite{kim2018measuring}} & \multirow{3}*{Ethereum} & Network size and geographic distribution of Ethereum network nodes & Proposed a network monitoring tool named NodeFinder, which is designed to find the unusual network properties of Ethereum network nodes in the underlying P2P network perspective.\\
	 
	\hline
	\multirow{3}*{\cite{alsahan2020local}}  & \multirow{3}*{Bitcoin network} &  TPS, network latency, number of forks, and mining rewards & The authors proposed a local Bitcoin network simulator to study the performance of Bitcoin under different network conditions including various topologies, network latencies, packet loss rates, and mining difficulties.\\

	\hline
	
\end{tabular}
\label{Table:measurements}
\end{table*}

 Nasir \textit{et al.} \cite{nasir2018performance} conducted performance measurements and discussion of two versions of Hyperledger Fabric. The authors focused on the metrics including execution time, transaction latency, throughput and the scalability versus the number of nodes in blockchain platforms. Several useful insights have been revealed for the two versions of Hyperledger Fabric. 
 As already mentioned previously in \cite{wang2019monoxide},  the authors evaluated their proposed Monoxide w.r.t the metrics including the scalability of TPS as the number of network zones increase, the overhead of  both cross-zone transactions and storage size, the confirmation latency of transactions, and the orphan rate of blocks.
 In \cite{yang2019prism}, the authors performed rich measurements for their proposed new blockchain protocol Prism under limited network bandwidth and CPU resources. The  performance evaluated includes the distribution of block propagation delays, the relationship between block size and mining rate, block size versus assembly time, the expected time to reach consensus on block hash, the expected time to reach consensus on blocks, etc. 

 Later, Zheng \textit{et al.} \cite{zheng2018detailed} proposed a scalable framework for monitoring the real-time performance blockchain systems. This work has evaluated four popular blockchain systems, i.e., Ethereum, Parity \cite{parity}, Cryptape Inter-enterprise Trust Automation (CITA) \cite{cita} and Hyperledger Fabric \cite{androulaki2018hyperledger}, in terms of several metrics including \textit{transactions per second}, \textit{average response delay}, \textit{transactions per CPU}, \textit{transactions per memory second}, \textit{transactions per disk I/O} and \textit{transactions per network data}.
 Such comprehensive performance evaluation results offered us rich viewpoints on the 4 popular blockchain systems.
 Their experimental logs and technique report \cite{xblock2020performance} can be accessed from \url{http://xblock.pro}. 
 Recently, Zheng \textit{et al.} \cite{zheng2019xblock} extended their work and released a new open-source dataset framework, called XBlock-ETH, for the data-driven analysis of Ethereum. XBlock-ETH contains multiple types of Ethereum data such as transactions, smart contracts and tokens. Thus, researchers can extract and explore the data of Ethereum using XBlock-ETH. The authors first collected and cleaned the most recent on-chain dataset from Ethereum. Then, they presented how to perform basic exploration of these datasets to make them best. Like their previous work, those datasets and processing codes can be found from the webpage \textit{xblock.pro} aforementioned.
 In the other similar work \cite{zheng2020xblock} of the same team, authors proposed another new dataset framework dedicated to EOSIO, named XBlock-EOS, which also includes multiple types of rich on-chain/off-chain datasets such as transactions, blocks, smart contracts, internal/external EOS transfer events, tokens, accounts and resource management. To show how to utilize the proposed framework, the authors presented comprehensive statistics and explorations using those datasets, for example, blockchain analysis, smart contract analysis, and cryptocurrency analysis. Finally, this study also discussed future directions of XBlock-EOS in the topics including: i) data analysis based on off-chain data to provide off-chain user behavior for blockchain developers, ii) exploring new features of EOSIO data that are different from those of Ethereum, and iii) conducting a joint analysis of EOSIO with other blockchains.

 }
 
\begin{table*}[t]
\caption{Blockchain Dataset Frameworks and Evaluation Tools.}
\centering
\footnotesize
\begin{tabular}{|p{0.15\textwidth}|p{0.1\textwidth}|p{0.03\textwidth}|p{0.58\textwidth}|}%
\hline
\textbf{Recognition}&\textbf{Target} &\textbf{Ref.} &\textbf{Utilization}\\
\hline

	\multirow{3}*{XBlock-ETH} & \multirow{3}*{Ethereum} &  \multirow{3}*{\cite{zheng2019xblock}} &  Authors released a new open-source dataset framework for analysis of Ethereum, i.e., XBlock-ETH, which includes multiple types of Ethereum datasets such as transactions, smart contracts and tokens. \\
	 
	\hline
	\multirow{2}*{XBlock-EOS}  & \multirow{2}*{EOS} &  \multirow{2}*{\cite{zheng2020xblock}} & Authors proposed a new dataset framework dedicated to EOSIO, named XBlock-EOS, to show how to perform comprehensive statistics and exploration of EOSIO datasets.\\
	
	\hline
	\multirow{2}*{BlockSci}   &  {General blockchains}  &  \multirow{2}*{\cite{kalodner2017blocksci}} & Authors proposed an open-source software platform, named BlockSci, for the analysis of blockchains.\\

	\hline
	\multirow{2}*{Blockbench}   & General blockchains &  \multirow{2}*{\cite{dinh2017blockbench}}  & Authors proposed a benchmarking framework for measuring the data processing capability and performance of different layers of a blockchain system.\\

	\hline
	\multirow{2}*{NodeFinder}   & Etheruem nodes &  \multirow{2}*{\cite{kim2018measuring}}  & Authors proposed a measuring tool named NodeFinder, to investigate the opaque network characteristics of Ethereum network nodes.\\

	\hline
	Network simulator for Bitcoin  & \multirow{2}*{Bitcoin} & \multirow{2}*{\cite{alsahan2020local}}  & Authors proposed a configurable network simulator for the performance measurements of Bitcoin using lightweight virtualization technologies.\\

	\hline
	
\end{tabular}
\label{Table:datasetFramework}
\end{table*}


\subsection{{\color{black}Useful Evaluation Tools for Blockchains}}

{\color{black}

  Kalodner \textit{et al.} \cite{kalodner2017blocksci} proposed BlockSci, which is designed as an open-source software platform for blockchain analysis. Under the architecture of BlockSci, the raw blockchain data is parsed to produce the core blockchain data including transaction graph, indexes and scripts, which are then provided to the analysis library. Together with the auxiliary data including P2P data, price data and user tags, a client can either directly query or read through a Jupyter notebook interface.

  To evaluate the performance of private blockchains, Dinh \textit{et al.} \cite{dinh2017blockbench} proposed a benchmarking framework, named Blockbench, which can measure the data processing capability and the performance of various layers of a blockchain system. Using such Blockbench, the authors then performed detailed measurements and analysis of three blockchains, i.e., Ethereum, Parity and Hyperledger. The results disclosed some useful experiences of those three blockchain systems. For example, today's blockchains are not scalable w.r.t data processing workloads, and several bottlenecks should be considered while designing different layers of blockchain in the software engineering perspective.
 
  Ethereum has received enormous attention on the mining challenges, the analytics of smart contracts, and the management of block mining. However, not so many efforts have been spent on the information dissemination in the perspective of P2P networks. To fill this gap, Kim \textit{et al.} \cite{kim2018measuring} proposed a measuring tool named NodeFinder, which aims to discover the opaque network properties of Ethereum network nodes. Through a three-month long data collection on the P2P network, the authors analyzed and found several unprecedented differences of Ethereum network comparing with other popular P2P networks like BitTorrent, Bitcoin and Gnutella in terms of network size and geographic distribution.
   
 Recently, by exploiting lightweight virtualization technologies,  Alsahan \textit{et al.} \cite{alsahan2020local} developed a configurable network simulator for the performance measurements of Bitcoin. The proposed simulator allows users to configure diverse network conditions, such as blockchain network topology, link delays, and mining difficulties, to emulate the real-world operation environment. Using this simulator, experiments can be performed to measure Bitcoin network under various network conditions. It also supports conducting the tests of security attacks and point of failure simulations. The authors also made this simulator open-source on Github.

 }

\section{Open Issues and Future Directions}\label{sec:openissue}

In this section, we envision the open issues and promising directions for future studies.
 
\subsection{Performance-Improving Issues}
\subsubsection{Scalability Issues}

 Scalability is still a severe challenge for most of the blockchain systems.
 For example, the PBFT consensus protocols issue a \textit{O}($n^{2}$) number of messages, where $n$ is the number of participants. The large number of messages makes the scalability unrealistic.
 Therefore, new distributed practical byzantine protocols and theoretical modelings of scalability solutions, such as sidechain, subchain, off-chain, sharding technique,  DAG, and even chain-less proposals, are in an urgent need for scalable blockchains.

\subsubsection{ Resilient Mechanisms for Sharding Technique }
  
  The sharding technique includes three typical categories, i.e., transaction sharding, network sharding, and state sharding. Via the extensive review on the existing studies of sharding techniques, we found that the resilient mechanisms for sharding blockchains are still missing. Particularly to the state sharding, once the failures occurred on blockchain nodes, how to ensure the correct recovery of the real-time running states in the failed blockchain  node(s) is critical to the resilience and robustness of the blockchain.

\subsubsection{ Cross-Shard Performance }
  
  Although a number of committee-based sharding protocols \cite{miller2016honey, kokoris2018omniledger, zamani2018rapidchain, wang2019monoxide} have been proposed, those protocols can only endure at most 1/3 adversaries. Thus, more robust byzantine agreement protocols need to be devised. Furthermore, all the sharding-based protocols incur additional cross-shard traffics and latencies because of the cross-shard transactions. Therefore, the cross-shard performance in terms of throughput, latency and other metrics, has to be well guaranteed in future studies.
  On the other hand, the cross-shard transactions are inherent for the cross-shard protocols. Thus, the pros and cons of such the correlation between different shards are worthy investigating using certain modelings and theories such as graph-based analysis.

\subsubsection{ Cross-Chain Transaction Accelerating Mechanisms }

 On cross-chain operations, \cite{jin2018towards} is essentially a pioneer step towards practical blockchain-based ecosystems. Following this roadmap paved by \cite{jin2018towards}, we are exciting to anticipate the subsequent related investigations will appear soon in the near future.
 For example, although the inter-chain transaction experiments achieve an initial success, we believe that the secure cross-chain transaction accelerating mechanisms are still on the way.
 In addition, further improvements are still required for the interoperability among multiple blockchains, such as decentralized load balancing smart contracts for sharded blockchains.

\subsubsection{ Ordering Blocks for Multiple-Chain Protocols }

Although multiple-chain techniques can improve the throughput by exploiting the parallel mining of multiple chain instances, how to construct and manage the blocks in all chains in a globally consistent order is still a challenge to the multiple-chain based scalability protocols and solutions.

\subsubsection{Hardware-assisted Accelerating Solutions for Blockchain Networks}
To improve the performance of blockchains, for example, to reduce the latency of transaction confirmation, some advanced network technologies, such as RDMA (Remote Direct Memory Access) and high-speed network cards, can be exploited in accelerating the data-access among miners in blockchain networks.

\subsubsection{Performance Optimization in Different Blockchain Network Layers}

The blockchain network is built over the P2P networks, which include several typical layers, such as mac layer, routing layer, network layer, and application layer. The BFT-based protocols are essentially working for the network layer. In fact, performance improvements can be achieved by proposing various protocols, algorithms, and theoretical models for other layers of the blockchain network.

\subsubsection{Blockchain-assisted BigData Networks}
Although big data and blockchain have several performance metrics that are contrary to each other. For example, big data is a centralized management technology with an emphasize on the privacy-preserving oriented to diverse computing environments. The data processed by big data technology should ensure nonredundancy and unstructured architecture in a large-scale computing network. In contrast, blockchain technology builds on a decentralized, transparent and immutable architecture, in which data type is simple, data is structured and highly redundant. Furthermore, the performance of blockchains require scalability and the off-chain computing paradigm.
Thus, how to integrate those two technologies together and pursue the mutual benefit for each other is an open issue that is worthy in-depth studies. For example, the potential research topics include how to design a suitable new blockchain architecture for big data technologies, and how to break the isolated data islands using blockchains while guaranteeing the privacy issues of big data.

\subsection{Issues for Better Understanding Blockchains Further}

{\color{blue}

Although the state-of-the-art studies have reviewed a lot of modelings and theories for better understanding blockchains, more sophisticated approaches and insightful mechanisms are still needed to help researchers gain a new level of perception over the high-performance blockchain systems.}
Some interesting directions are summarized here for inspiring more subsequent investigations.

\begin{itemize}
\item Exploiting more general queueing theories to capture the real-world arrival process of transactions, mining new blocks, and other queueing-related blockchain phases.

\item Performing priority-based service policies while dealing with transactions and new blocks, to meet a predefined security or regulation level.

\item Developing more general probabilistic models to characterize the correlations among the multiple performance parameters of blockchain systems.
\end{itemize}

\subsection{\modified{Security Issues of Blockchains}}

 \subsubsection{Privacy-Preserving for Blockchains} 
 
 From the previous overview, we observe that most of the existing works under this category are discussing the blockchain-based security and privacy-preserving applications. The fact is that the security and privacy are also the critical issues of the blockchain itself. For example, the privacy of transactions could be hacked by attackers. However, dedicated studies focusing on those issues are still insufficient.

\subsubsection{Anti-Cryptojacking Mechanisms for Malicious Miners}

The Cryptojacking Miners are reportedly existing in web browsers according to \cite{tahir2019browsers}. This type of malicious codes is commandeering the hardware resources such as computational capability and memory of web users. Thus, the anti-cryptojacking mechanisms and strategies are necessary to develop for protecting normal browser users.

\subsubsection{Security Issues of  Cryptocurrency Blockchains}

 The security issues of cryptocurrency blockchains, such as double-spend attacks, frauds in smart contracts, have arisen growing attention from both industrial and academic fields. However, little efforts have been committed to the theoretical investigations towards the security issues of cryptocurrency blockchains. For example, the exploration of punishment and cooperation between miners over multiple chains is an interesting topic for cryptocurrency blockchains. Thus, we expect to see broader perspectives of modeling the behaviors of both attackers and counterattackers in the context of monetary blockchain attacks.

\subsection{Powerful Experimental Platforms for Blockchains}

 To most of the beginners in the field of the blockchain, they have a dilemma about lack of powerful simulation/emulation tools for verifying their new ideas or protocols. Therefore, the powerful simulation/emulation platforms that are easy to deploy scalable testbeds for the experiments would be very helpful to the research community.


\section{Conclusion}\label{sec:conclusion} 

Through a brief review of state-of-the-art blockchain surveys at first, we found that a dedicated survey focusing on the theoretical modelings, analytical models and useful experiment tools for blockchains is still missing.
To fill this gap, we then conducted a comprehensive survey of the state-of-the-art on blockchains, particularly in the perspectives of theories, modelings, and measurement/evaluation tools.
The taxonomy of each topic presented in this survey tried to convey the new protocols, ideas, and solutions that can improve the performance of blockchains, and help people better understand the blockchains in a further level.
We believe our survey provides a timely guidance on the theoretical insights of blockchains for researchers, engineers, educators, and generalized readers.

\section{Acknowledgement}

This work was supported in part by the Key-Area Research and Development Program of Guangdong Province (No. 2019B020214006), the National Natural Science Foundation of China (No. 61902445, No. 61722214, No. 61872310), the Fundamental Research Funds for the Central Universities of China (No.19lgpy222), the Guangdong Basic and Applied Basic Research Foundation (No. 2019A1515011798), the Hong Kong RGC Research Impact Fund (RIF)  (No. R5060-19, No. R5034-18), General Research Fund (GRF) (No. 152221/19E), and the Collaborative Research Fund (CRF) (No. C5026-18G).

\bibliographystyle{IEEEtran}
\bibliography{reference}

\end{document}